\documentclass[widetext,showpacs,preprintnumbers,superscriptaddress]{revtex4}

\usepackage[dvipdfm]{graphicx}
\usepackage{amssymb}
\usepackage{amsmath}
\usepackage{dcolumn}
\usepackage{bm}
\usepackage{natbib}
\usepackage{multirow}
\usepackage{aas_macros}

\begin{document}

\newcommand{\rem}[1]{{\bf #1}}

\preprint{ICRR-Report-577-2010-10,\ IPMU10-0221,\ YITP-10-110}

\title{Evolution of String-Wall Networks and Axionic Domain Wall Problem}

\author{Takashi Hiramatsu}
\email{hiramatz@yukawa.kyoto-u.ac.jp}
\affiliation{Yukawa Institute for Theoretical Physics, Kyoto University, 
Kitashirakawa Oiwake-cho, Sakyo-ku, Kyoto 606-8502, Japan}
\author{Masahiro Kawasaki}
\email{kawasaki@icrr.u-tokyo.ac.jp}
\affiliation{Institute for Cosmic Ray Research, The University of Tokyo, 
5-1-5 Kashiwa-no-ha, Kashiwa City, Chiba 277-8582, Japan} 
\affiliation{Institute for Physics and Mathematics of the Universe, 
The University of Tokyo,
5-1-5 Kashiwa-no-ha, Kashiwa City, Chiba 277-8582, Japan}
 \author{Ken'ichi Saikawa}
  \email{saikawa@icrr.u-tokyo.ac.jp}
   \affiliation{Institute for Cosmic Ray Research, The University of Tokyo, 
5-1-5 Kashiwa-no-ha, Kashiwa City, Chiba 277-8582, Japan} 
 
\date{\today}

\begin{abstract}
We study the cosmological evolution of domain walls bounded by strings which arise naturally in axion models. 
If we introduce a bias in the potential, walls become metastable and finally disappear. 
We perform two dimensional lattice simulations of domain wall networks and estimate the decay rate of domain walls. 
By using the numerical results, we give a constraint for the bias parameter and the Peccei-Quinn scale. 
We also discuss the possibility to probe axion models by direct detection of gravitational waves produced by domain walls.
\end{abstract}


\maketitle

\section{\label{sec1}Introduction}
In the history of the early universe, various types of phase transitions have presumably occurred as a consequence
of the spontaneous breaking of symmetries in fundamental physics. During this process, some nontrivial vacuum structures,
called topological defects, would be formed, and they can affect the evolution of the universe. For example,
the formation of cosmic strings has rich phenomenological consequences for cosmology such as the structure formations~\cite{1994csot.book.....V}.
On the other hand, domain walls, which arise when a discrete symmetry become spontaneously broken, are considered as
cosmologically undesirable objects~\cite{1974JETP...40....1Z}.

These topological defects arise in theories beyond the standard model of particle physics. A well-known example is the theory of axions~\cite{2010RvMP...82..557K}. 
Axion~\cite{1978PhRvL..40..223W, 1978PhRvL..40..279W} is the pseudo-Nambu-Goldstone boson which arise as a consequence of the Peccei-Quinn (PQ)
mechanisms~\cite{1977PhRvL..38.1440P, 1977PhRvD..16.1791P} which is introduced to solve
the CP violation problem in quantum chromodynamics (QCD). In this model, one can make the CP violating phase dynamically into zero value by introducing a new 
symmetry called U(1)$_{\mathrm{PQ}}$. In the early universe, this symmetry has been spontaneously broken when the temperature of the universe has fallen below
some energy scale (the PQ scale). This is called the PQ phase transition. At this time the networks of global strings called axionic strings are formed. 
Furthermore, at the time of the QCD phase transition, the axion acquires a mass, which causes the spontaneous breaking of discrete
$Z_{N_{\mathrm{DW}}}$ subgroup of U(1)$_{\mathrm{PQ}}$, leading to the formation of domain walls attached to the strings.
The integer parameter $N_{\mathrm{DW}}$ describes the number of degenerate vacua and the number of walls attached to the string.
The value of this number depends on the models. For example, in the Dine-Fischler-Srednicki-Zhitnitsky (DFSZ)
models~\cite{1981PhLB..104..199D, Zhitnitsky1980}, $N_{\mathrm{DW}}=2N_g$, where $N_g$ is the number of quark generations (this can be reduced to
$N_{\mathrm{DW}}=N_g$, depending on the structure of the Higgs sector~\cite{1990PhRvD..41.3848G}).
On the other hand, in the Kim-Shifman-Vainshtein-Zakharov (KSVZ) models~\cite{1979PhRvL..43..103K, 1980NuPhB.166..493S},
one can get $N_{\mathrm{DW}}=1$.

There are two possibilities for the fate of such string-wall networks. One is the case with $N_{\mathrm{DW}}=1$.
In this case, walls bounded by strings quickly slice themselves into small peaces as discussed in \cite{1982PhRvL..48.1867V, 1984PhRvD..30.2036V}.
Another possibility is the case with $N_{\mathrm{DW}}>1$. In this case the networks survive for a long time, which is cosmologically
disastrous~\cite{1982PhRvL..48.1156S} since they overclose the universe and distort the cosmic microwave
background (CMB) observed today~\cite{1974JETP...40....1Z}. Therefore, axion models with $N_{\mathrm{DW}}>1$ seem to conflict with the standard cosmology.
This is called the axionic domain wall problem.

There are several ways to avoid this problem. The simplest way is just to consider the model with $N_{\mathrm{DW}}=1$. In this model, the walls decay immediately
after the QCD phase transition, producing a radiation of barely relativistic axions \cite{1999PhRvD..59b3505C}. This can be realized in the KSVZ models.
Also, it is possible to take $N_{\mathrm{DW}}=1$ in the variant axion models~\cite{1986PhLB..172..435P, 1986PhLB..173..189K},
and its collider implications are recently investigated in~\cite{2010JHEP...06..059C}.  A more intricate solution is to embed the discrete subgroup
$Z_{N_{\mathrm{DW}}}$ of U(1)$_{\mathrm{PQ}}$ in the center of another continuous group
(so called the Lazarides-Shafi mechanism~\cite{1982PhLB..115...21L,1987PhR...150....1K}). In this model, the degenerate vacua are connected to each other by
another symmetry transformation. However, in this kind of model, one have to choose the symmetry group,
Higgs representations and U(1)$_{\mathrm{PQ}}$ charge judiciously, which seems to be unlikely to occur~\cite{1982PhLB..116..227B}.
The third possibility is to suppose that inflation has occurred after the PQ phase transition to dilute the number density of strings as well as domain walls.
In this case, however, a constraint which comes from the bound on isocurvature perturbations in CMB observation might be severe~\cite{2008PhRvD..78h3507H}.
Alternatively, one can eliminate domain walls by introducing a tiny $Z_{N_{\mathrm{DW}}}$ breaking term (i.e. a bias) which lifts the vacuum
degeneracy \cite{1982PhRvL..48.1156S, 1999PhRvD..59b3505C, 2008LNP...741...19S}. In this case, domain walls collapse due to the pressure force acting
between different vacua~\cite{1981PhRvD..23..852V}.

In this paper we mainly consider the last possibility, biased domain walls, as a solution to the domain wall problem.
It is natural to expect such a explicit $Z_{N_{\mathrm{DW}}}$ breaking terms in the context of the quantum theory of gravity such as string theory:
The quantum gravitational effects induce higher-dimensional operators suppressed by powers of the Planck mass, which may alter the
PQ solution to the CP violating problem~\cite{1992PhLB..282..137K, 1992PhLB..282..132H, 1992PhRvD..46..539B, 1997PhRvD..55.5826D}.
It is argued that allowed region in the parameter space for this solution is narrow, though it is not ruled out~\cite{2008LNP...741...19S}.

The cosmological scenario in which domain walls decay at early time is also interesting from the observational point of view.
It has been pointed out that the emission of gravitational waves is likely to occur due to the decay of domain walls~\cite{1999PhRvD..59b3505C}.
Indeed, the recent numerical study implies that long-lived unstable domain walls can radiate gravitational waves with the amplitude
enough to observe~\cite{2010JCAP...05..032H}. If it is true, it might be possible to probe axion models by gravitational wave experiments in the next decades.

We would like to address these issues more quantitatively, based on the field theory lattice simulations. The numerical simulations of axionic domain walls
have been performed by several authors. The simulation of decaying domain wall with $N_{\mathrm{DW}}=1$ was
examined in~\cite{1999PhRvD..59b3505C, 1993PhLB..318...53N, 1994PhRvD..50.4821N},
neglecting the expansion of the universe. Also, the evolution of string-wall networks with $N_{\mathrm{DW}}\ge1$ models in the expanding background
was investigated in~\cite{1990ApJ...357..293R}. In addition to these simulations, we investigate the decay process of the networks due to the explicit
$Z_{N_{\mathrm{DW}}}$ breaking term, give the numerical confirmation of the decay of the networks with $N_{\mathrm{DW}}>1$,
estimate their lifetime, and constraint the parameter space in which the CP violation problem and the domain wall problem are solved simultaneously.

This paper is organized as follows. In section 2, we describe the model which we consider, and briefly review the property of string-wall networks.
In section 3, we present the results of the numerical simulations and estimate the decay time of the networks.
Based on these results, we give observational constraints for the bias parameter and the PQ scale in section 4.
In section 5, we shortly discuss about a possibility to observe gravitational waves from domain wall networks in the future experiments.
Then we conclude in section 6.

{\bf Notation}: we work in the spatially flat Friedmann-Robertson-Walker (FRW) background in which the metric is given by
\begin{equation}
ds^2 = dt^2 - a^2(t)[dx^2+dy^2+dz^2]. \nonumber
\end{equation}
We denote the cosmic time as $t$ and the conformal time as $\tau$, where $d\tau = dt/a(t)$.

\section{\label{sec2}Dynamics of domain wall networks with multiple vacua}
We consider the model of a complex scalar field (the PQ field) $\phi$ with the Lagrangian density given by
\begin{equation}
{\cal L} = \frac{1}{2}\partial_{\mu}\phi^*\partial^{\mu}\phi - V(\phi), \label{eq2-1}
\end{equation}
and the potential given by
\begin{equation}
V(\phi) = \frac{\lambda}{4}(\phi^*\phi - \eta^2)^2 + \frac{m^2\eta^2}{N_{\mathrm{DW}}^2}(1-\cos N_{\mathrm{DW}}\theta) + \delta V, \label{eq2-2}
\end{equation}
where $\theta$ is the phase of $\phi$ (i.e. $\phi=|\phi| e^{i\theta}$), $m$ corresponds to the mass of the axion, and $\eta$ corresponds to the PQ scale.
The first term in eq.~(\ref{eq2-2}) describes the usual mexican hat potential which causes the spontaneous breaking
of U(1)$_{\mathrm{PQ}}$ and produces cosmic strings.
Since we are interested in the evolution of domain walls produced after the QCD phase transition, we must include the effective potential for the axion
field, which is represented as the second term in eq.~(\ref{eq2-2}). In the absence of the last term in eq.~(\ref{eq2-2}),
the theory has $Z_{N_{\mathrm{DW}}}$ shift symmetry $\theta \to \theta + 2\pi k/N_{\mathrm{DW}}\ (k=0,1,\dots, N_{\mathrm{DW}}-1$).
The spontaneous breaking of this symmetry implies the occurrence of $N_{\mathrm{DW}}$ domain walls attached to the string.
However, we include the additional term in eq.~(\ref{eq2-2}) which explicitly breaks the discrete
$Z_{N_{\mathrm{DW}}}$ symmetry~\cite{1982PhRvL..48.1156S,2008LNP...741...19S}:
\begin{equation}
\delta V = -\xi\eta^3(\phi e^{-i\delta} + \mathrm{h.c.}). \label{eq2-3}
\end{equation}
Due to this term, domain walls become unstable and eventually decay into the true vacuum at which $|\delta - \theta|$ is smallest.
This kind of term may arise from some ultimate theories, but we ignore the detail and simply describe their effect as 
a dimensionless parameter $\xi$ (for now, $\delta$ and $\xi$ are regarded as free parameters).

First we consider the case with $\xi=0$. If we restrict ourselves to low energy configurations at the QCD scale, 
we can put $\phi=\eta e^{i\theta}$ and derive the effective Lagrangian for $\theta$
\begin{equation}
{\cal L}_{\mathrm{eff}} = \frac{\eta^2}{2}(\partial_{\mu}\theta)^2 + \frac{m^2\eta^2}{N_{\mathrm{DW}}^2}(1-\cos N_{\mathrm{DW}}\theta). \label{eq2-4}
\end{equation}
The equation of motion derived from it in the Minkowski background has the sine-Gordon solution which describes a planer wall
orthogonal to the $z$-axis~\cite{1994csot.book.....V}
\begin{equation}
\theta(z) = \frac{2\pi k}{N_{\mathrm{DW}}} + \frac{4}{N_{\mathrm{DW}}}\tan^{-1}\exp(mz),\qquad k=0,1,\dots, N_{\mathrm{DW}}-1. \label{eq2-5}
\end{equation}
From this equation we can estimate the thickness of the wall as $\delta_w\simeq m^{-1}$. The surface mass density of the wall is also estimated as
$\sigma = 8m\eta^2/N_{\mathrm{DW}}^2$ if we use this exact solution. However, if we include the structure of the neutral pion field which
varies inside the wall, the surface energy density of the wall becomes a slightly larger value~\cite{1985PhRvD..32.1560H}
\begin{equation}
\sigma \simeq 4.3f_{\pi}m_{\pi}\eta/N_{\mathrm{DW}} \simeq 9m\eta^2/N_{\mathrm{DW}}^2, \label{eq2-6}
\end{equation}
where $f_{\pi}$ is the pion decay constant and $m_{\pi}$ is the mass of the pion.

By using the exact solution (\ref{eq2-5}), we can also estimate the gradient energy stored in the domain wall
\begin{equation}
\rho_{\mathrm{grad}} = \frac{1}{2}|\nabla\phi|^2 = \frac{\eta^2}{2}\left(\frac{d\theta}{dz}\right)^2 =\frac{2\eta^2m^2}{N_{\mathrm{DW}}^2}\frac{1}{\cosh^2(mz)}. \label{eq2-7}
\end{equation}
Note that it is equal to the height of the potential energy $2m^2\eta^2/N_{\mathrm{DW}}^2$ at the center of the wall ($z=0$). Therefore, if there are planer walls,
their gradient energy becomes comparable with their potential energy, as we see in the next section.

It has been argued that the evolution of domain wall networks is characterized by one scale, the Hubble radius, and 
the averaged number of walls per Hubble volume remain the same in the evolution of the universe.
Such a property is called the scaling solution, and confirmed both
numerically~\cite{1989ApJ...347..590P, 2003PhRvD..68j3506G, 2005PhLB..610....1A, 2005PhRvD..71h3509O}
and analytically~\cite{1996PhRvL..77.4495H, 2003PhRvD..68d3510H, 2005PhRvD..72h3506A}
for a simple model in which the domain walls arise from the spontaneous breaking of $Z_2$ symmetry.
Also, the numerical simulation performed in \cite{1990ApJ...357..293R} indicates that this property is true for
networks of $N_{\mathrm{DW}}$ domain walls attached to strings unless $N_{\mathrm{DW}}=1$. If domain wall networks
are in the scaling regime, the energy density of domain walls evolves as
\begin{equation}
\rho_{\mathrm{walls}} \sim \sigma/t. \label{eq2-8}
\end{equation}
This is equivalent to the fact
\begin{equation}
A/V \propto \tau^{-1}, \label{eq2-9}
\end{equation}
where $A/V$ is the comoving area density occupied by domain walls. We will check this property in the numerical simulation described in the next section.

If we turn on the bias ($\xi\ne 0$), walls become unstable and finally disappear. Let us estimate the typical time scale for the annihilation of walls.
The energy difference between the neighboring vacua introduced by the $Z_{N_{\mathrm{DW}}}$ breaking term (\ref{eq2-3}) can be estimated as
$\Delta V \sim \frac{2\pi}{N_{\mathrm{DW}}}2\xi\eta^4$, which acts as a volume pressure $p_V$ on the wall and accelerates it against the false vacuum regions:
\begin{equation}
p_V \sim \Delta V \sim 4\pi\xi\eta^4/N_{\mathrm{DW}}. \label{eq2-10}
\end{equation}
On the other hand, the surface tension which straightens the wall up to the horizon scale can be estimated as
\begin{equation}
p_T \sim \sigma /t. \label{eq2-11}
\end{equation}
The domain walls collapse when these two effects become comparable. From this fact, we can estimate the typical time of the decay of domain wall networks as
\begin{equation}
t_{\mathrm{dec}} \sim \sigma/\Delta V \sim \frac{2m}{\pi\xi N_{\mathrm{DW}}\eta^2}, \label{eq2-12}
\end{equation}
where we used eq.~(\ref{eq2-6}) for $\sigma$. We will determine the precise value of the numerical coefficient in the formula~(\ref{eq2-12}) from
the results of the numerical simulations in the next section.

\section{\label{sec3}Lattice simulations}
The dynamical equations which describe the evolution of the PQ field are highly nonlinear. 
Therefore, we must perform numerical simulations in order to obtain any quantitative result.
In this section, we describe the results of the numerical simulations.
First, we present the evolution of unbiased domain walls (i.e. $\xi=0$) and check the consistency of the result with other existing numerical simulations.
Next, we investigate the $\xi$ dependence of the results and estimate the lifetime of domain wall networks. We also comment on the $m$ dependence,
which is the dependence on the ratio between the axion mass and the PQ scale which we take as a free parameter in the numerical simulations.
The details for the numerical computations are summarized in appendix~\ref{secA}.

\subsection{\label{sec3-1}Evolution of the stable domain walls}
We solve the classical field equations derived from the Lagrangian (\ref{eq2-1}) [see eqs.~(\ref{eqA-1}) and (\ref{eqA-2})] on the two dimensional lattice with $4096^2$ points.
We used two dimensional simulations since they enable us to investigate the evolution of networks with long dynamical range.
Memory of the computer limits us to much smaller grid size and shorter dynamical range, if we perform three dimensional simulations.
In the earlier work about the evolution of the string-wall networks~\cite{1990ApJ...357..293R},
it was confirmed that there are no significant discrepancy between the results of two dimensional and three dimensional simulations.

We choose the value of the parameters as $\lambda=0.1$ and $m/\eta=0.1$. Note that $m/\eta$ represents the ratio between the axion mass
and the PQ scale. This value for $m/\eta$ is much larger than
the actual value of $m/\eta$ by many orders of magnitude. It is impossible to use the actual value of $m/\eta$ in the lattice simulations since $m^{-1}$ represents
the width of the domain wall, which must be smaller than the box size. Therefore we perform the simulations with $m/\eta=0.1$ and extrapolate the results
into the actual value of $m/\eta$. We will mention this point again in section~\ref{sec3-2-2}.

For stable domain walls, we performed the simulations varying the value of $N_{\mathrm{DW}}$ from 1 to 6. The final time of the simulation
is set to be $\tau_f=110$ in the unit of $\eta^{-1}$. Even at this time, the resolution of the wall width and the core of stings is well maintained and the Hubble radius is smaller
than the simulation box (see appendix~\ref{secA1}).

\subsubsection{\label{sec3-1-1}Field configurations and time evolution of the area density}
\begin{figure}[htbp]
\begin{center}
\setlength{\tabcolsep}{3pt}
\begin{tabular}{l l}
\resizebox{83mm}{!}{\includegraphics{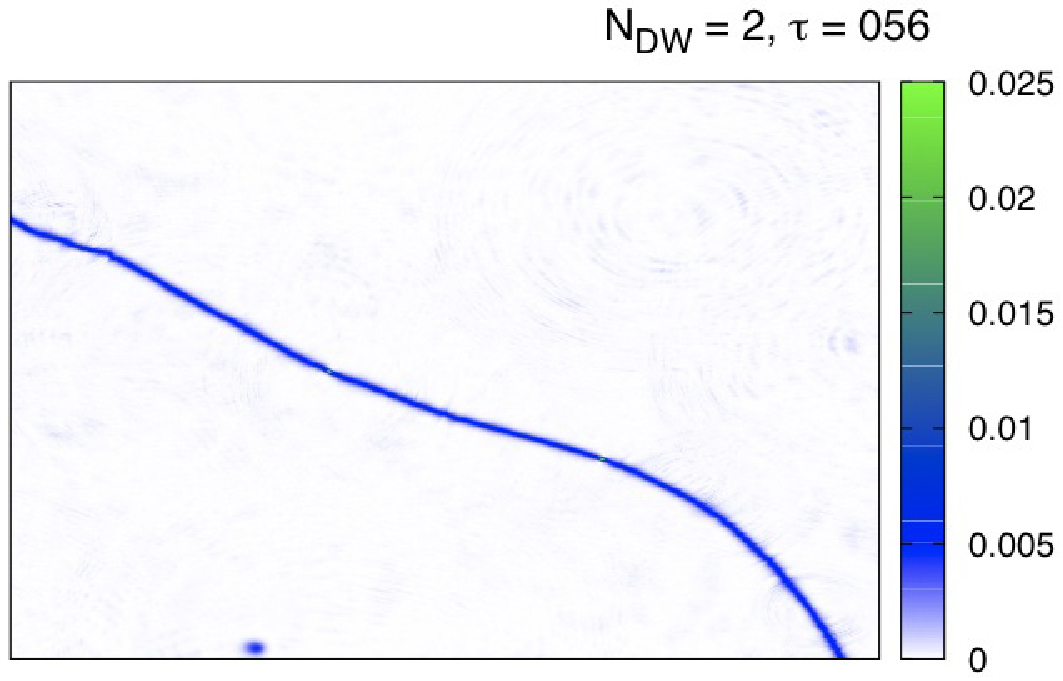}} &
\resizebox{83mm}{!}{\includegraphics{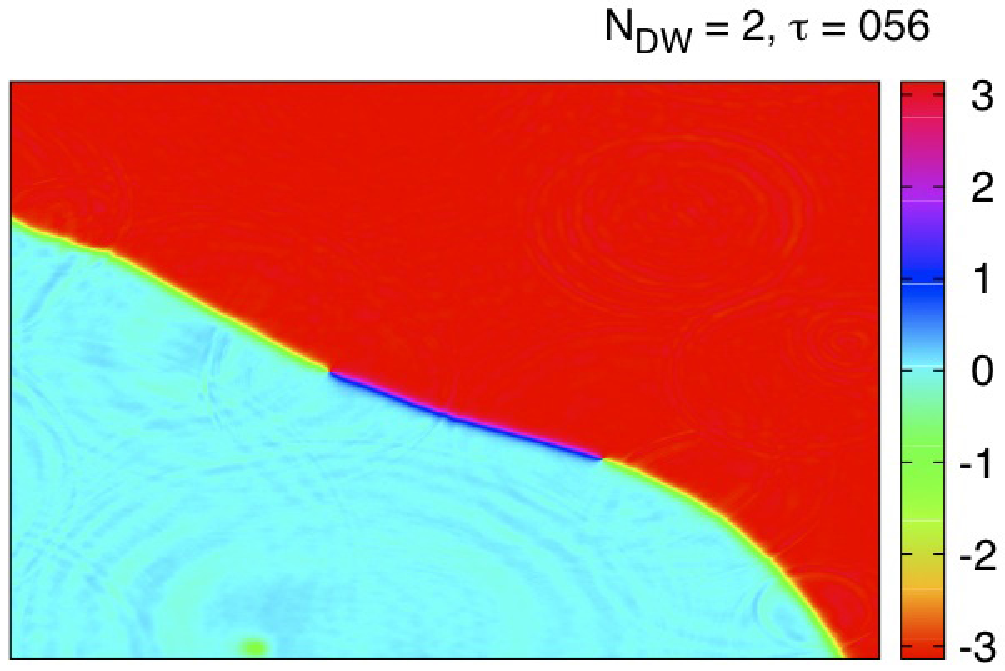}} \\
\resizebox{83mm}{!}{\includegraphics{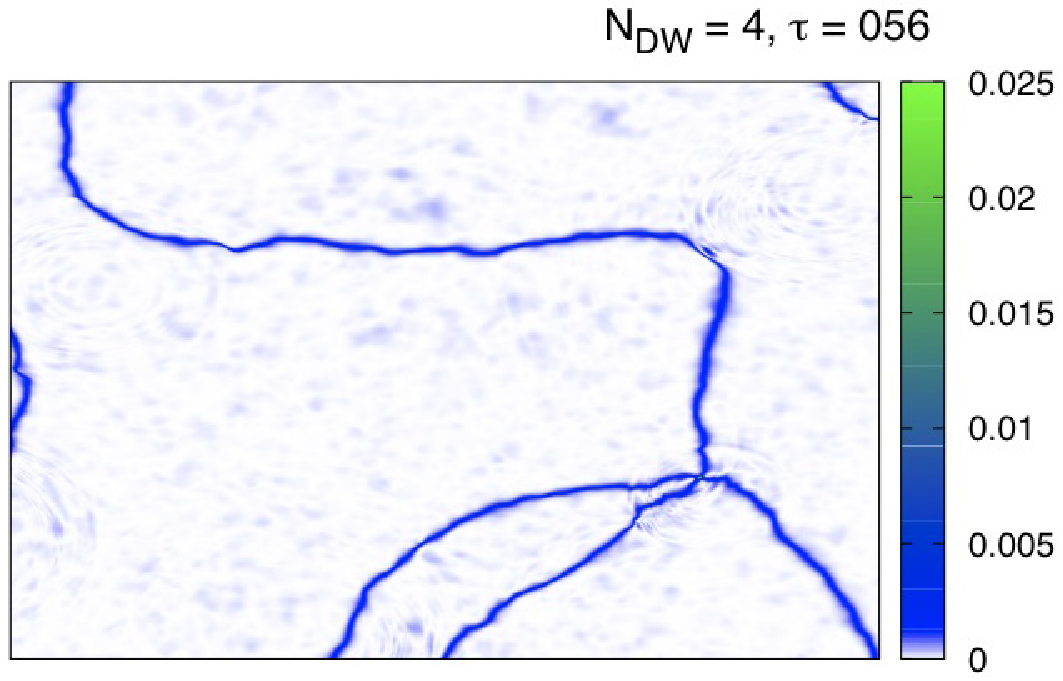}} &
\resizebox{83mm}{!}{\includegraphics{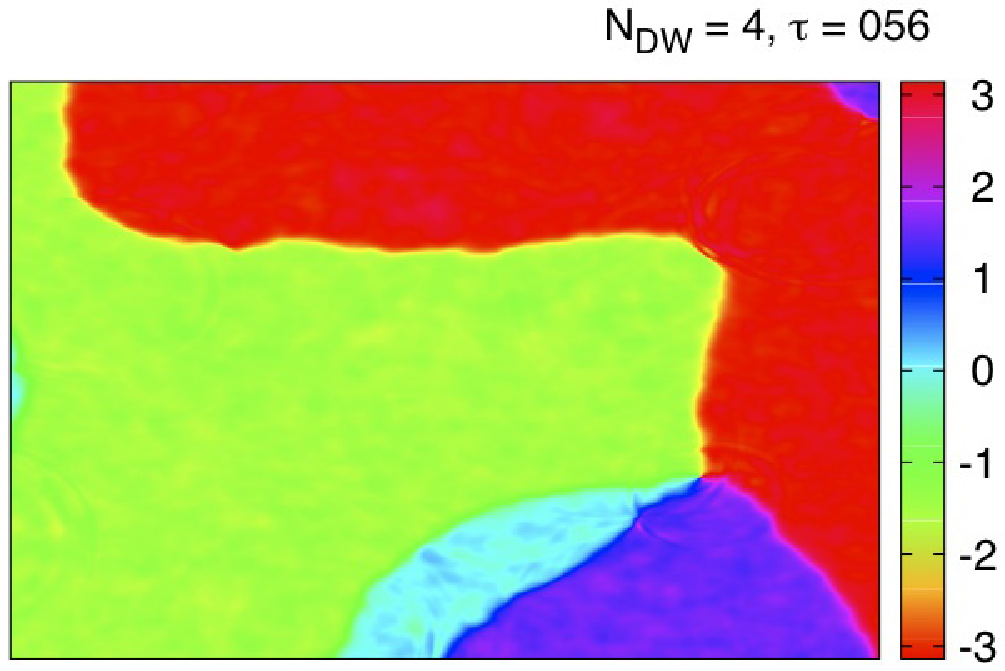}} \\
\resizebox{83mm}{!}{\includegraphics{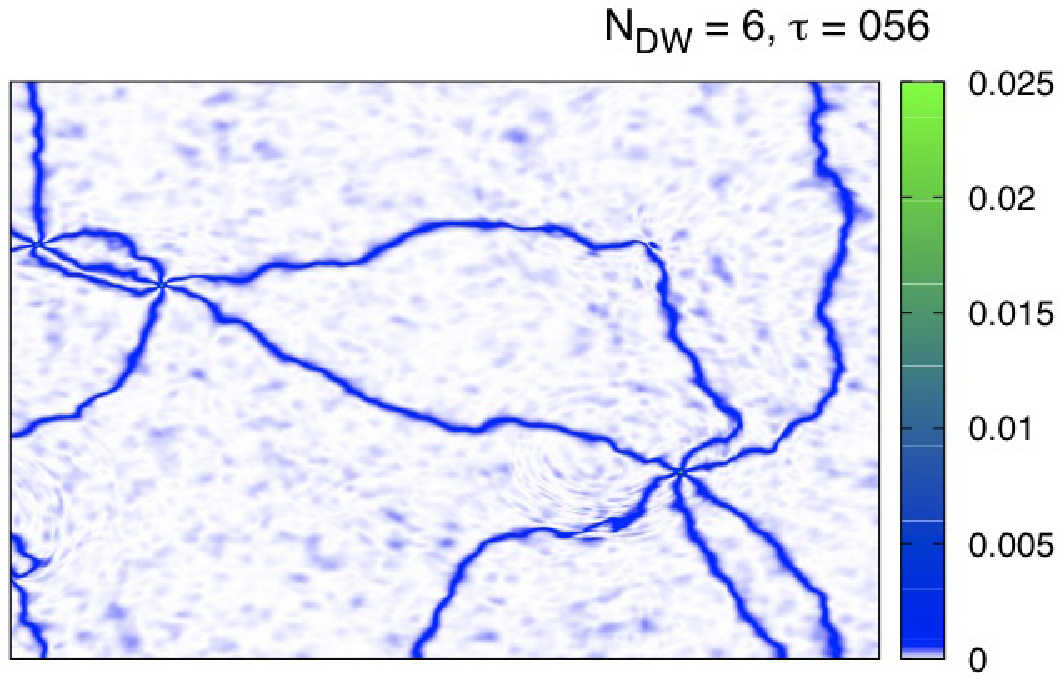}} &
\resizebox{83mm}{!}{\includegraphics{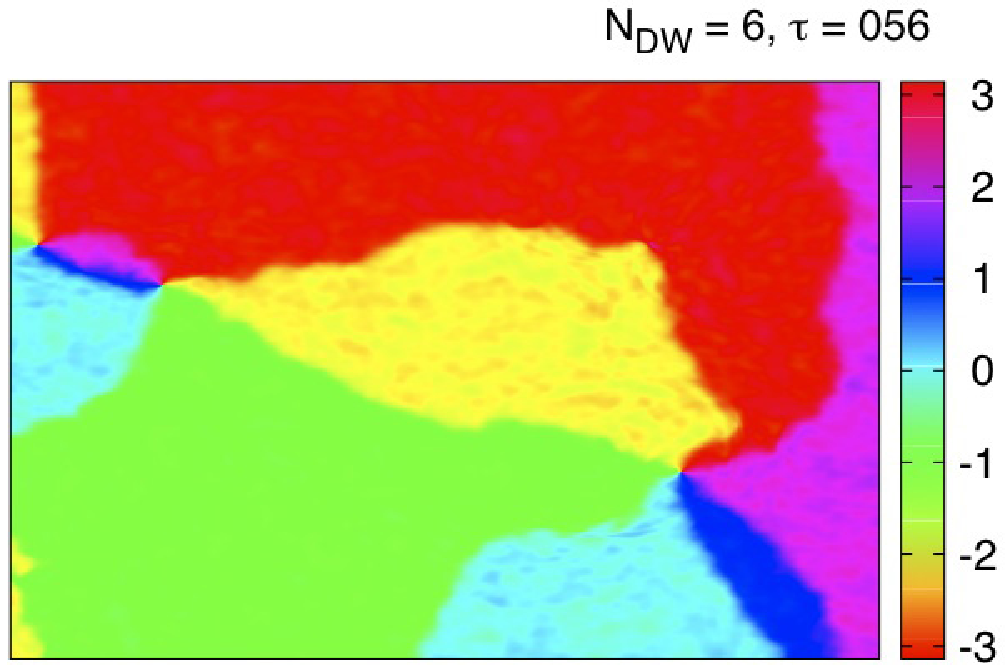}} \\
\end{tabular}
\end{center}
\caption{The distributions of the potential energy (left) and the phase of the PQ field $\theta$ (right) with the case $N_{\mathrm{DW}}=2$ (top), $N_{\mathrm{DW}}=4$ (middle), and $N_{\mathrm{DW}}=6$ (bottom) at the time $\tau=56$. The size of these figures is set to be a quarter of the size of the simulation box. In the distribution of the energy density, the white region corresponds to the vacuum ($V(\phi)\simeq0$), the blue region corresponds to the domain wall ($V(\phi)\simeq2m^2\eta^2/N_{\mathrm{DW}}^2$), and the green region corresponds to the string ($V(\phi)\simeq\lambda\eta^4/4$), but the core of the string is too thin to see in this figure. We take the range of $\theta$ as $-\pi<\theta<\pi$ in the right panel.}
\label{fig1}
\end{figure}

The spatial distribution of the potential energy and the phase of the PQ field are shown in figure \ref{fig1}. We can see that the phase of the PQ field is indeed divided into
$N_{\mathrm{DW}}$ domains and there is the core of strings attached by $N_{\mathrm{DW}}$ domain walls at the location where $\theta$ rotates by $2\pi$.
Since these are the results of two dimensional simulations, strings exist as a ``point" in the two dimensional surface.

\begin{figure}[htbp]
\begin{center}
\includegraphics{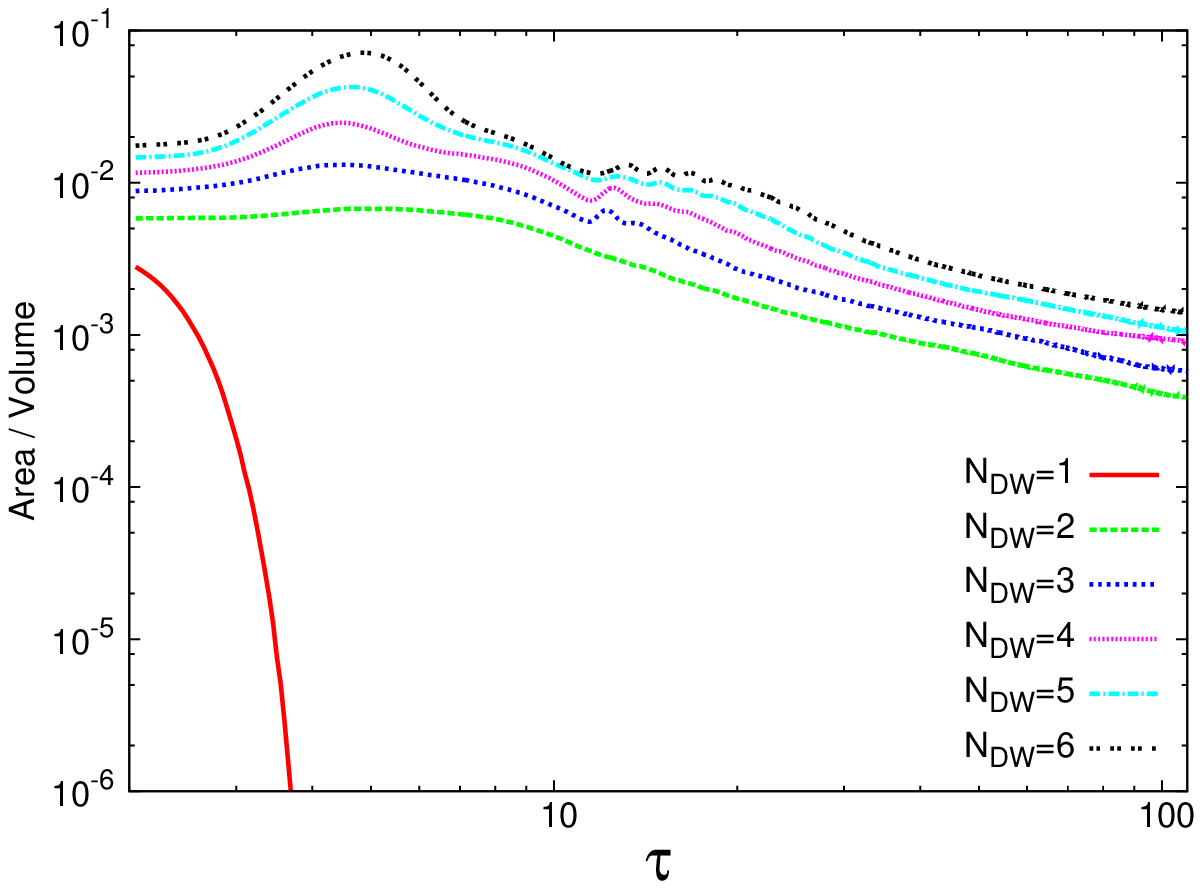}
\end{center}
\caption{The time evolution of the comoving area density of domain walls for various values of $N_{\mathrm{DW}}$.}
\label{fig2}
\end{figure}

We calculated the time evolution of the comoving area density of domain walls as shown in figure~\ref{fig2}. In the case with $N_{\mathrm{DW}}>1$,
the scaling regime in which the area density scales as $\tau^{-1}$ [see eq.~(\ref{eq2-9})] begins around $\tau\simeq10$. On the other hand, if $N_{\mathrm{DW}}=1$,
the area density falls off shortly after the beginning of the simulation. These properties are similar to that found in the past simulations~\cite{1990ApJ...357..293R}.

\subsubsection{\label{sec3-1-2}Time evolution of the energy density}
\begin{figure}[htbp]
\setlength{\tabcolsep}{20pt}
\begin{tabular}{c c}
\resizebox{75mm}{!}{\includegraphics{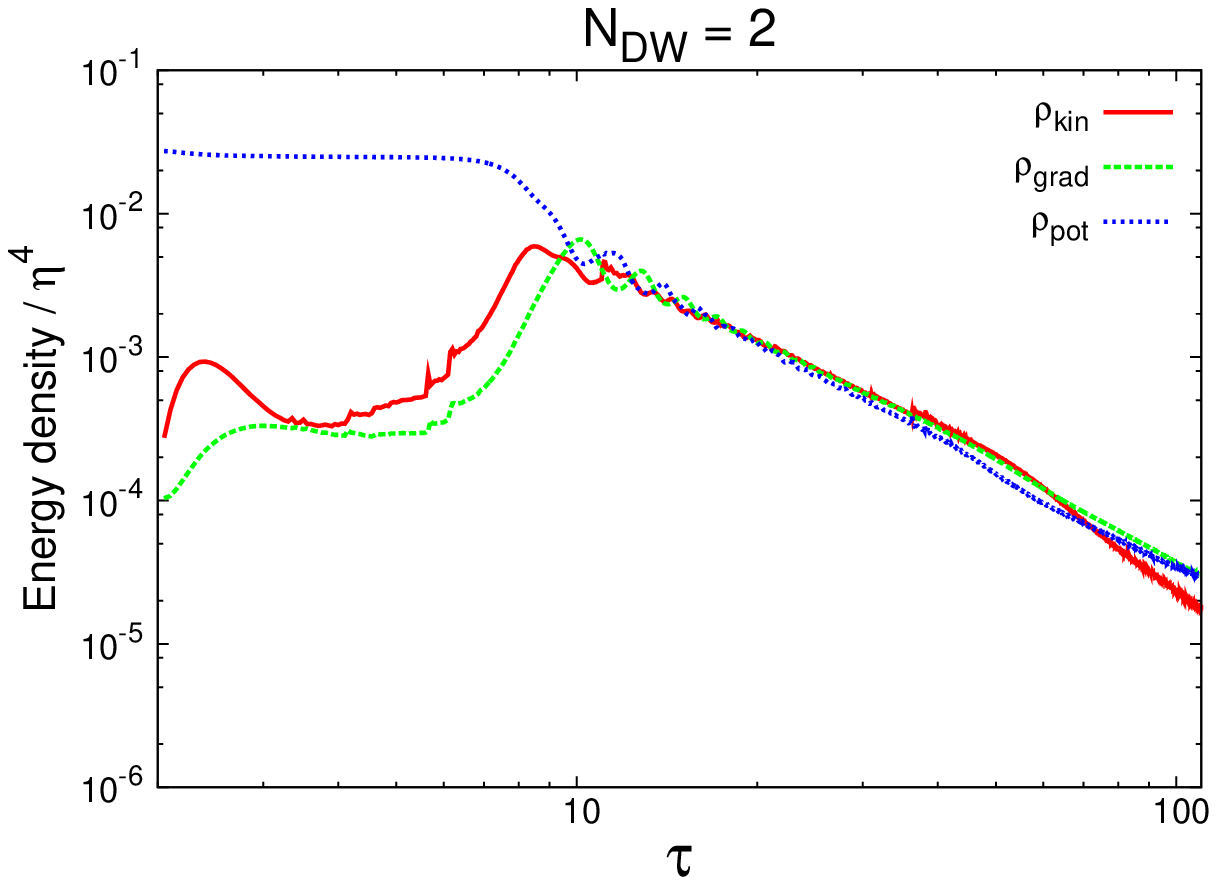}} &
\resizebox{75mm}{!}{\includegraphics{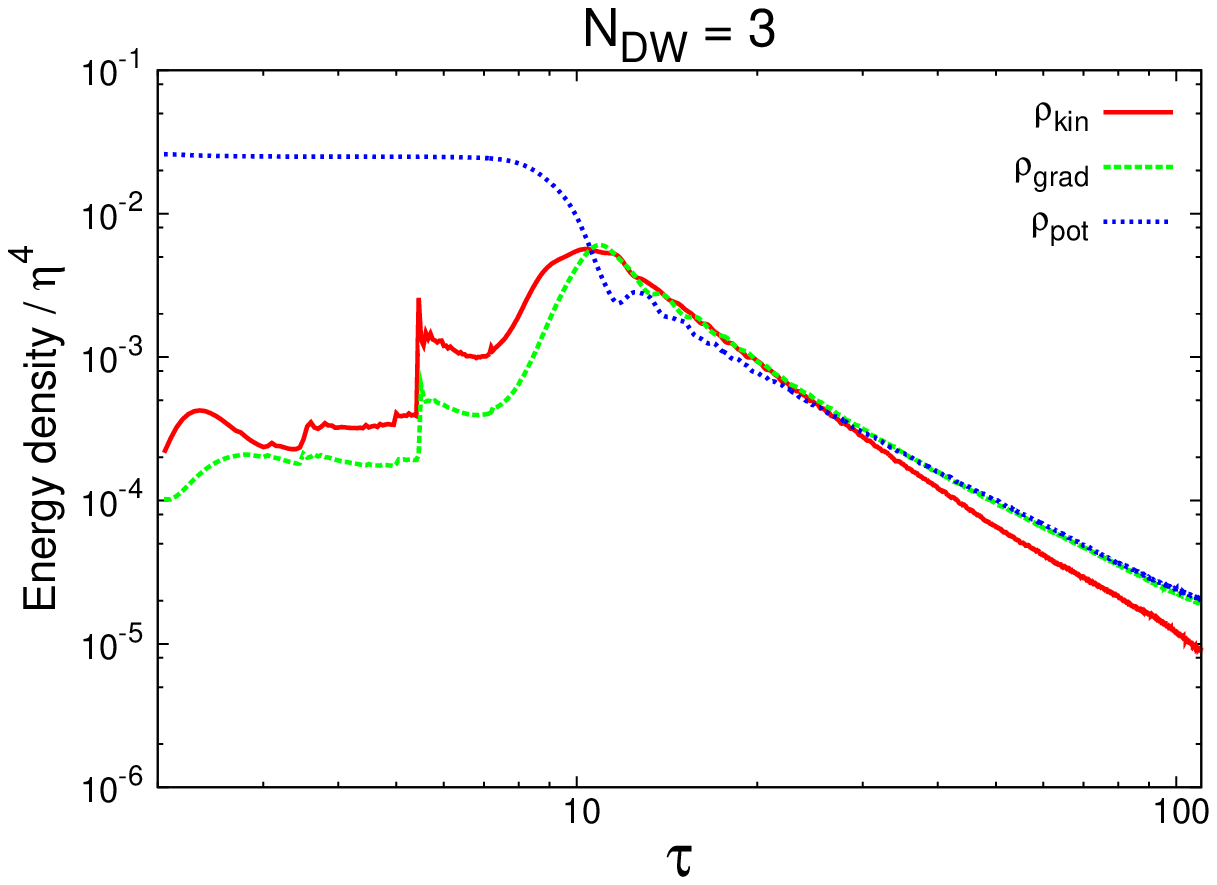}} \\
\resizebox{75mm}{!}{\includegraphics{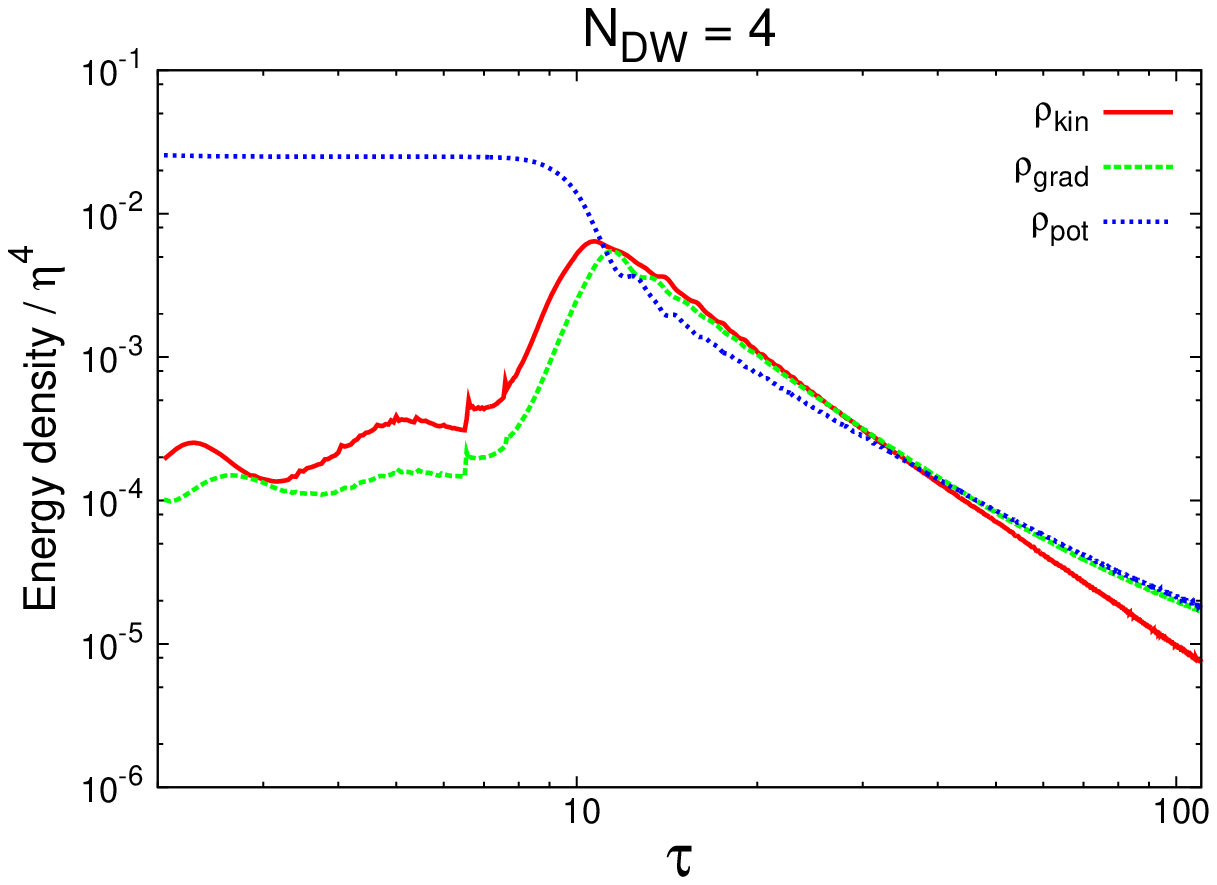}} &
\resizebox{75mm}{!}{\includegraphics{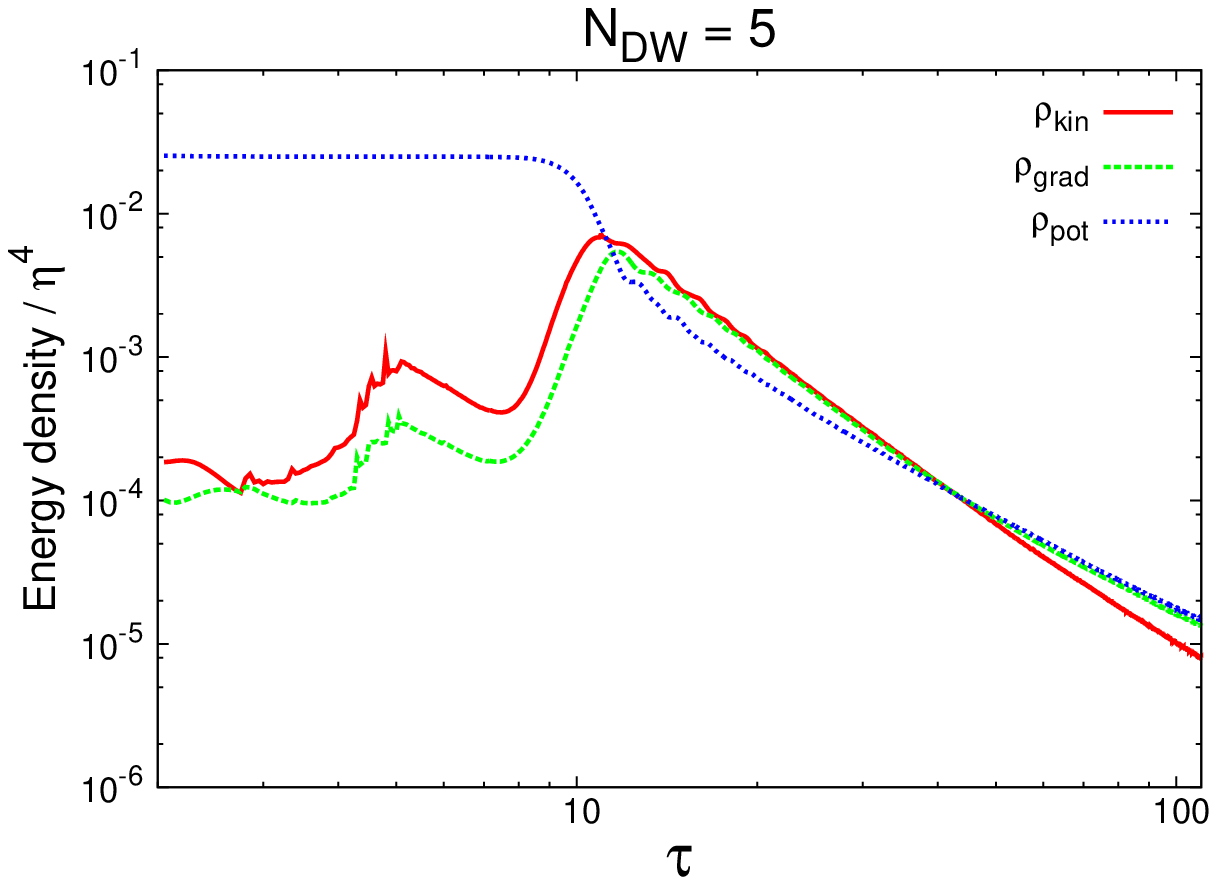}} \\
\resizebox{75mm}{!}{\includegraphics{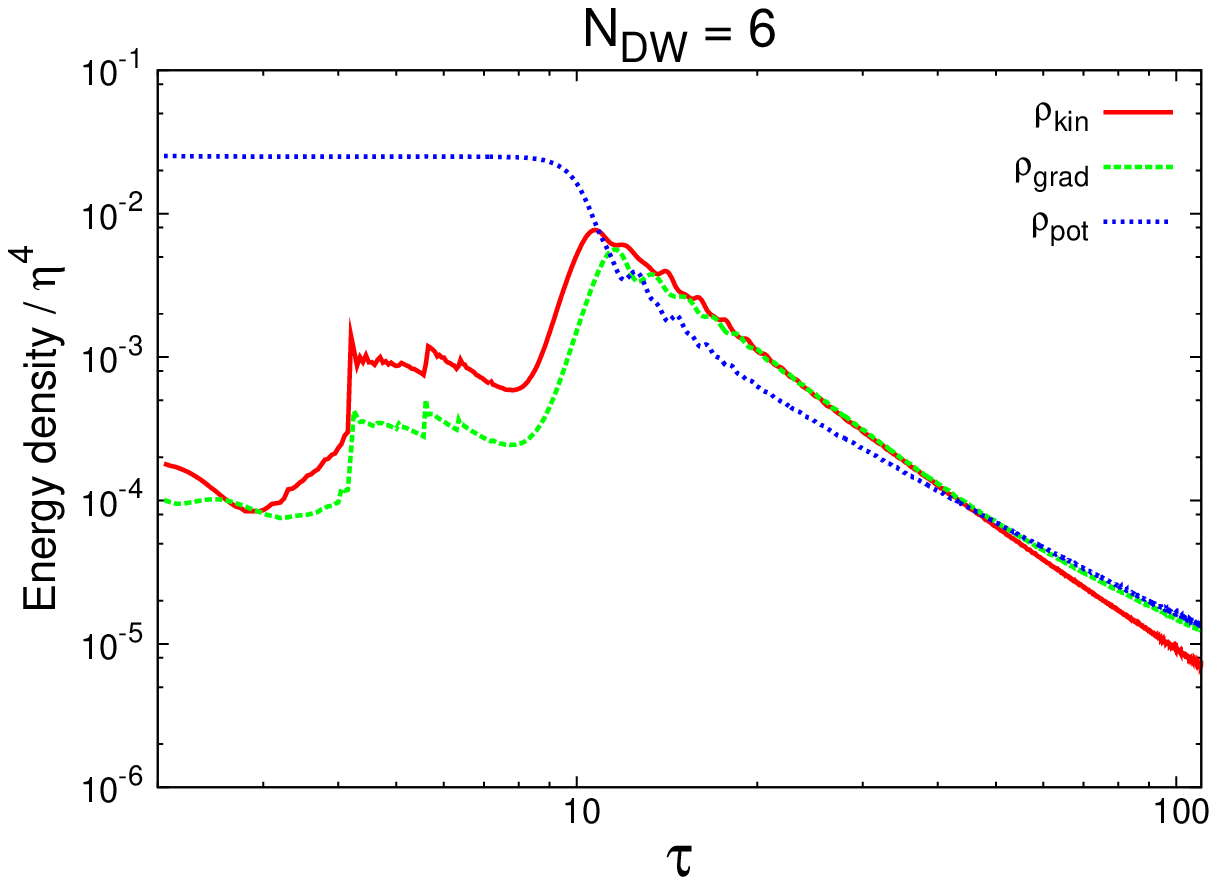}} & \\
\end{tabular}
\caption{The time evolution of the kinetic/gradient/potential energy densities of domain walls in the unit of $\eta^4$ for various values of $N_{\mathrm{DW}}$.}
\label{fig3}
\end{figure}

We also calculated the kinetic, gradient, and potential energy densities of the PQ field as shown in figure~\ref{fig3}. 
From this figure, we can see that the gradient and potential energy densities evolve as $t^{-1}\propto \tau^{-2}$ [cf. eq.~(\ref{eq2-8})] from the onset of the scaling regime
around $\tau\simeq10$. However, the kinetic energy decays faster than the gradient and potential energies. This is caused by
the cosmological expansion that dilutes the energy density of the ``radiation" component of the PQ field (the oscillation around the minimum of the potential)
which redshifts as $a^{-4}(t)$. On the other hand, the potential energy and the gradient energy become comparable after the formation of
scaling domain walls as we have shown in section~\ref{sec2}.
But the gradient energy is slightly larger than the potential energy at the early stage of the evolution, especially in the result with 
large $N_{\mathrm{DW}}$. It might be caused by the fact that if $N_{\mathrm{DW}}$ is large, there are many nodes in the networks.
In this case, walls are bent in small scale and the gradient energy stored in walls become larger than the potential energy.
As domain walls straighten up to the horizon scale, the gradient energy becomes small and eventually comes up with the potential energy.

We find some discontinuities in the plots of figure~\ref{fig3}. We regard these discontinuities as numerical artifacts rather than physical features.
The reason is as follows: We note that there is a term proportional to $1/|\phi|$ in the equation of motion of the scalar field [see eqs.~(\ref{eqA-1}) and (\ref{eqA-2})].
This term becomes singular when the scalar field is on the top of the mexican hat potential ($|\phi|=0$) and causes a sudden transition of the field values.
This might produce the spiky signatures in the plots of the kinetic and gradient energy densities of the scalar field.
On the other hand, since the potential energy does not contain the term proportional to $1/|\phi|$ [see eq.~(\ref{eqA-3})],
there are no discontinuities in the plot of the potential energy density.
In any case, it seems that this singularity does not affect the evolution of the field after the formation of domain walls since the factor $1/|\phi|$ has a definite value $1/\eta$.
We performed simulations in different setups (such as the simulation with the time variable chosen to be cosmic time rather than conformal time and the simulations
with different spatial resolution), and found that although the behavior of the kinetic and gradient energies in each setup seems to be different,
it does not affect the evolution after the formation of scaling domain walls.
Therefore, we expect that these numerical artifacts hardly alter the final result about the estimation of the decay time of domain wall networks.

In the past numerical simulation performed by Ryden et al.~\cite{1990ApJ...357..293R}, it was insisted that the ratio between
the kinetic energy and the total energy of the PQ field tends to be approximately constant and it does not strongly depend on the number $N_{\mathrm{DW}}$.
However, we can not confirm this fact in our simulations because of the difference in the dynamical range:
We take the dynamical range as $\simeq \tau_f/\tau_i =55$ in conformal time,
but in~\cite{1990ApJ...357..293R} the dynamical range was taken as $\simeq 200$.
This is because the simulation performed in~\cite{1990ApJ...357..293R} relied on
the modified equation of motion which keeps the resolution of the comoving wall width constant in time to improve the dynamical range of the simulation
(called PRS algorithm~\cite{1989ApJ...347..590P}). Now we do not use this formalism since it may erase small-scale structure on the wall~\cite{1994csot.book.....V},
which might affect the evolution of networks. Therefore, even if the ratio between the kinetic and total energies tends to be constant value,
it is not seen in our simulation time scale.

\subsection{\label{sec3-2}Evolution of unstable domain walls and estimation of the decay time}
Next we consider the effect of the term which explicitly breaks the discrete symmetry [i.e. eq.~(\ref{eq2-3})]. This effect is parameterized by
two quantities: $\delta$ and $\xi$. In general, $\delta$ can be defined relative to the phase in the quark mass matrix and have any value,
but in the numerical simulations it only determines the location of the true minimum in the potential and does not affect the results of the simulations.
Therefore, we take $\delta=0$ for a simplicity.

We note that the lift of the degenerate vacua must be sufficiently small since we assume the circumstance in which
the discrete symmetry is held approximately. We can understand this requirement more quantitatively as follows.
If the lift of the degenerate vacua is large, the probability of choosing vacuum at the time of formation of domain walls is not uniform
between different vacua. For example, assume that there are two vacua ($N_{\mathrm{DW}}=2$), and the 
energy density of one vacuum (false vacuum) is greater than that of another vacuum (true vacuum) due to the presence of the bias term $\delta V$.
Then, let us denote the probability of having a scalar field fluctuation at the time of the formation of domain walls
end up in true vacuum as $p_t$ and in false vacuum as $p_f$. The ratio of these two probabilities is given by~\cite{2010JCAP...05..032H}
\begin{equation}
\frac{p_f}{p_t} = \exp\left(-\frac{\Delta V_{\mathrm{bias}}}{\Delta V_{\mathrm{pot}}}\right) \equiv \exp(-R), \label{eq3-2-1}
\end{equation}
where $\Delta V_{\mathrm{pot}} \simeq 2m^2\eta^2/N_{\mathrm{DW}}^2$ is the height of the axion potential, and
$\Delta V_{\mathrm{bias}}\simeq 2\xi\eta^4$ is the difference of the energy densities between two vacua. These probabilities are
not uniform (i.e. $p_t=p_f=0.5$) if $\xi\ne 0$. It was shown that this non-uniform initial probability distribution also
destabilizes domain walls~\cite{1996PhRvD..53.4237C,1997PhRvD..55.5129L}. Let us define the parameter $\epsilon$
which represents the deviation from the uniform probability distribution as $p_t=0.5+\epsilon$. According to the numerical simulations~\cite{1997PhRvD..55.5129L},
the time scale of the decay of domain walls $\tau_{\mathrm{dec,prob}}$ (in conformal time) due to this effect is given by
\begin{equation}
\tau_{\mathrm{dec,prob}}/\tau_{\mathrm{form}} \simeq \epsilon^{-D/2}, \label{eq3-2-2}
\end{equation}
where $\tau_{\mathrm{form}}$ is the time of the formation of domain walls, and $D$ is the spatial dimension.
We must require that this time scale should be greater than the simulation time scale ($\tau_{\mathrm{dec,prob}}>\tau_f$),
since we would like to check the effect of volume pressure [i.e. the relation given by eq.~(\ref{eq2-12})] as a decay mechanism of domain walls.
In our numerical simulations, we choose $\tau_f=110$ and $\tau_{\mathrm{form}}\simeq 10$. Then the above requirement leads the condition
\begin{equation}
R = \frac{\xi\eta^2N_{\mathrm{DW}}^2}{m^2} < 0.4. \label{eq3-2-3}
\end{equation}
Note that this result is obtained for the case with $N_{\mathrm{DW}}=2$. It is not so straightforward to generalize above arguments for the case with multiple vacua.
However, we guess that the condition for the case with $N_{\mathrm{DW}}>2$ might be less severe than that given by eq.~(\ref{eq3-2-3}), since
if $N_{\mathrm{DW}}$ is large, the relative probability of choosing true vacuum becomes small and less sensitive to $R$.
Therefore, we take eq.~(\ref{eq3-2-3}) as a criterion to ignore the effect of non-uniform initial probability distribution even for the case with $N_{\mathrm{DW}}>2$.
It is difficult to satisfy this requirement for large $N_{\mathrm{DW}}$ in the numerical simulation since the
height of the axion potential is proportional to $N_{\mathrm{DW}}^{-2}$ and we have to choose small value of $\xi$ to make $R$ satisfy
the condition (\ref{eq3-2-3}), which is forbidden because of the limitation of the dynamical range (domain walls can not decay in the simulation time scale).
This restriction forbids us to perform the simulations with $N_{\mathrm{DW}}=5$ and $6$.


\subsubsection{\label{sec3-2-1}Dependence on $\xi$ and $N_{\mathrm{DW}}$}
\begin{figure}[htbp]
\setlength{\tabcolsep}{20pt}
\begin{tabular}{c c}
\resizebox{75mm}{!}{\includegraphics{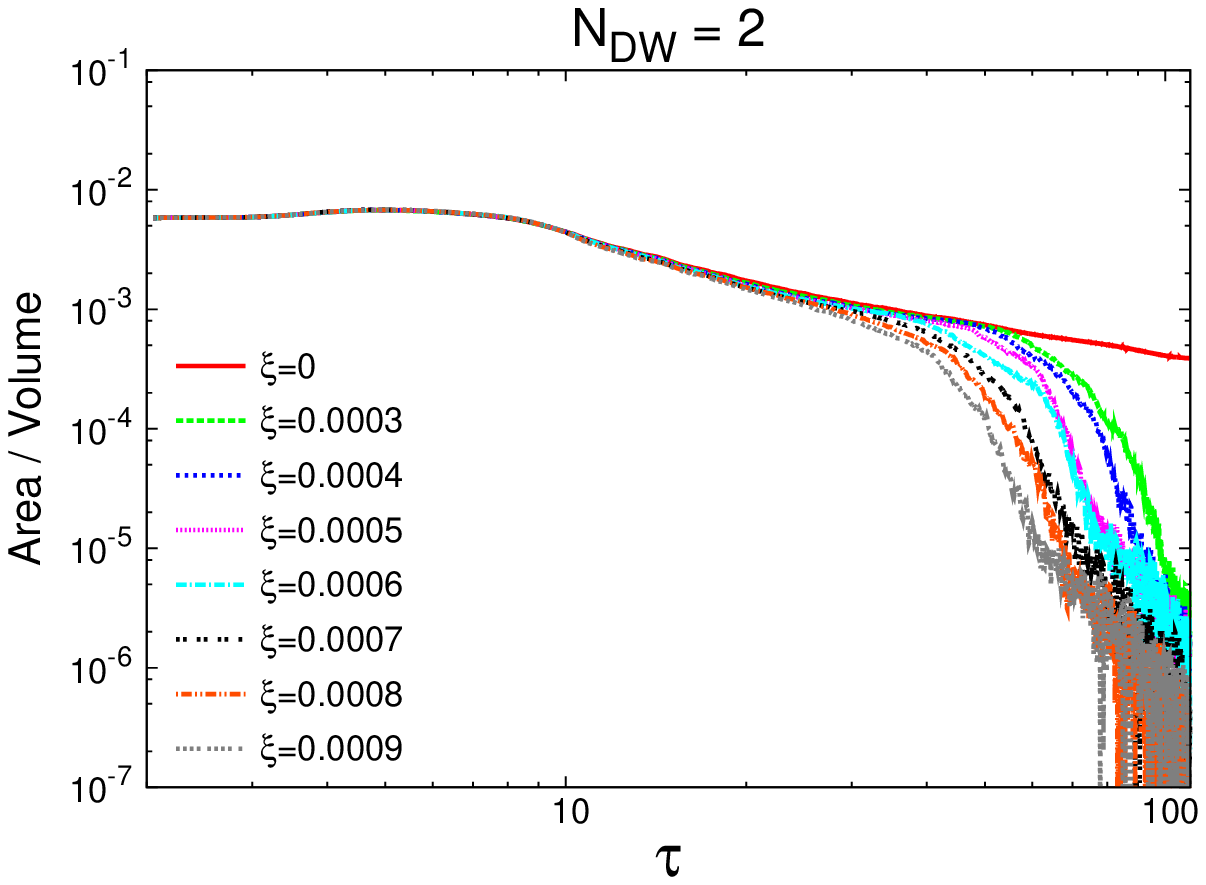}} &
\resizebox{75mm}{!}{\includegraphics{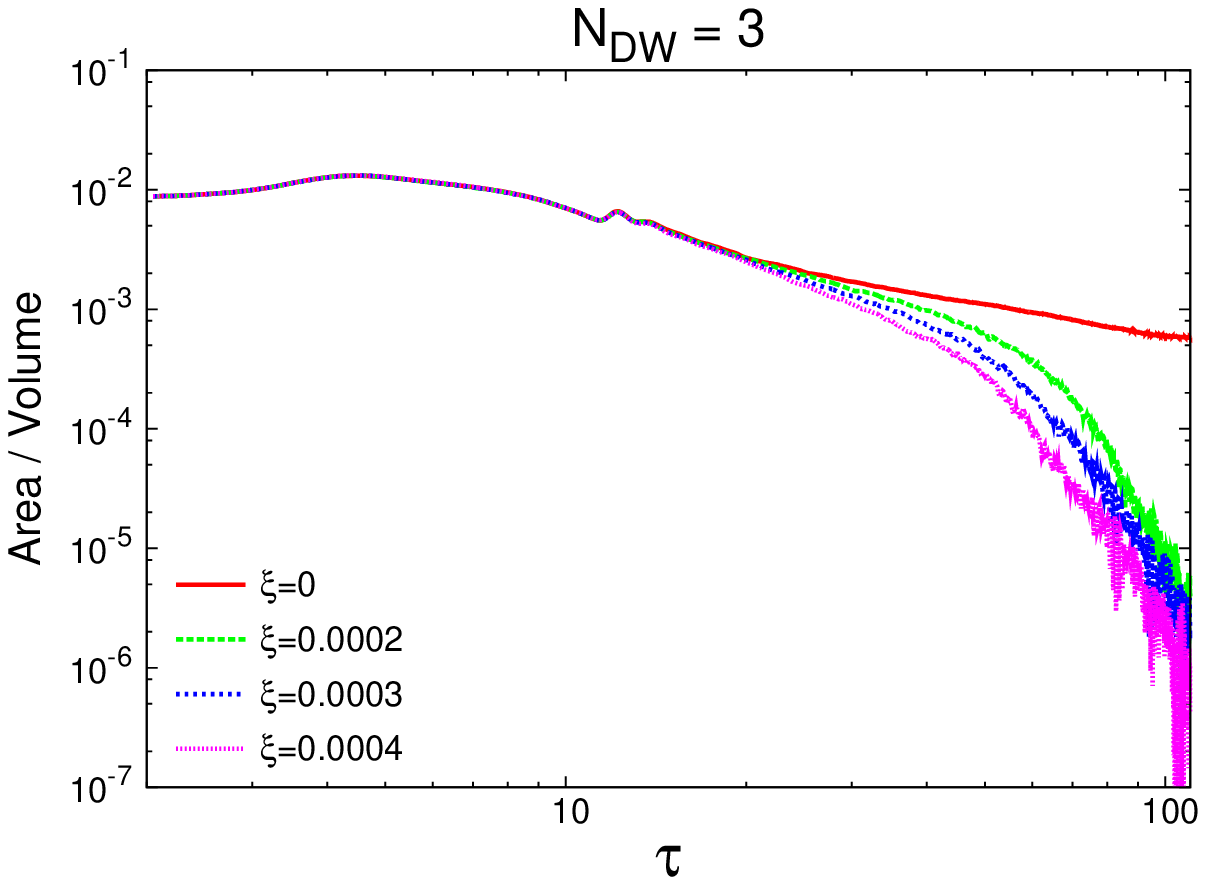}} \\
\resizebox{75mm}{!}{\includegraphics{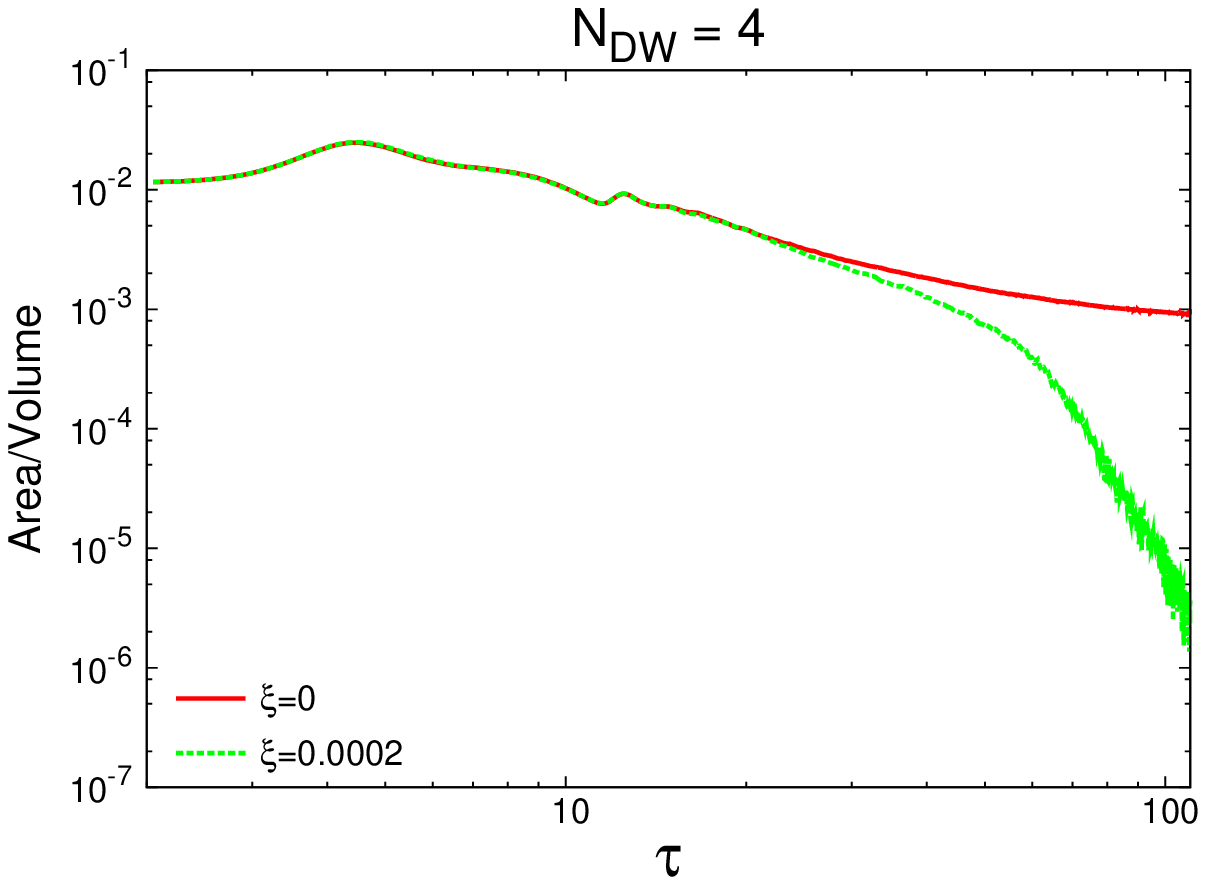}} & \\
\end{tabular}
\caption{The time evolution of the comoving area density of domain walls for various values of $\xi$ with $N_{\mathrm{DW}}=2$ (left top), $N_{\mathrm{DW}}=3$ (right top), and $N_{\mathrm{DW}}=4$ (bottom). In the case with $N_{\mathrm{DW}}=4$, we can choose only one parameter for $\xi$ because of the restriction given by eq.~(\ref{eq3-2-3}).}
\label{fig4}
\end{figure}

We performed the simulations for $N_{\mathrm{DW}}=2,3,4$, and $\xi$ satisfying the condition described above
and calculated the time evolution of the area density of domain walls. The results are shown in figure~\ref{fig4}.
We confirmed that domain walls decay in the time scale which we naively expect as eq.~(\ref{eq2-12}).

\begin{figure}[htbp]
\begin{center}
\includegraphics{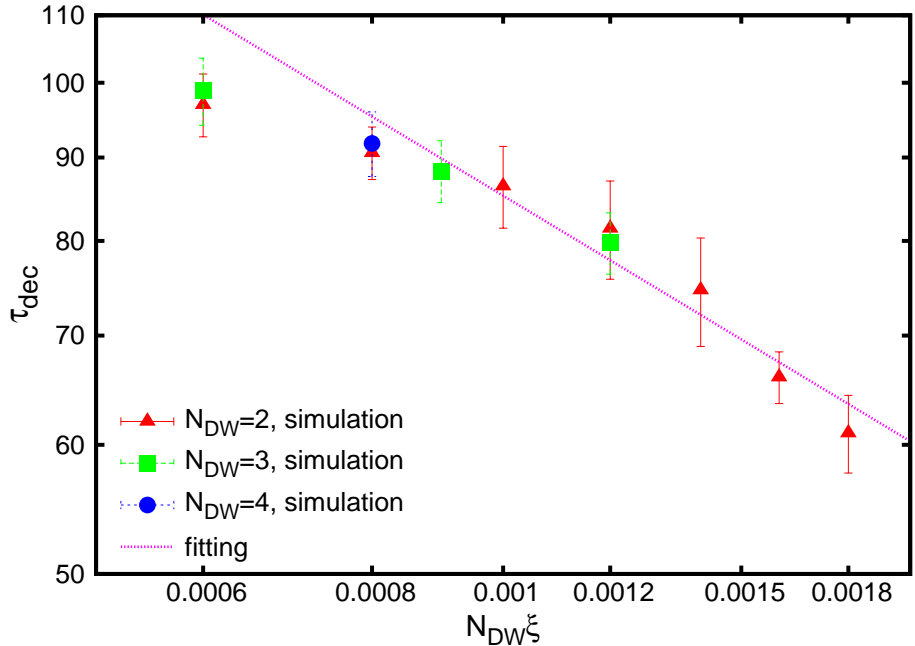}
\end{center}
\caption{The relation between $\tau_{\mathrm{dec}}$ and $N_{\mathrm{DW}}\xi$ obtained from the numerical results with $N_{\mathrm{DW}}=2$ (triangle), $N_{\mathrm{DW}}=3$ (square) and $N_{\mathrm{DW}}=4$ (circle). The dotted line represents the fitting given by eq.~(\ref{eq3-2-6}), which is obtained by using datas with $N_{\mathrm{DW}}\xi\ge0.001$.}
\label{fig5}
\end{figure}

Now we determine the time scale of the decay of the domain wall networks $\tau_{\mathrm{dec}}$ based on the results of the numerical simulations obtained here.
We define $\tau_{\mathrm{dec}}$ as a conformal time when the value of $A/V$ becomes 1\% of that with $\xi=0$. 
In figure~\ref{fig5}, we show the numerical results for $\tau_{\mathrm{dec}}$ averaging over 5 realizations.
We note that $\tau_{\mathrm{dec}}$ may depend on $N_{\mathrm{DW}}$ as we see in eq.~(\ref{eq2-12}).
Therefore, we plot the dependence on the product $N_{\mathrm{DW}}\xi$ in figure~\ref{fig5}.
We found that the results of the simulation with small $N_{\mathrm{DW}}\xi$ deviate from the simple scaling, but the results with large
$N_{\mathrm{DW}}\xi$ can be fitted to the naive analytic expectation (\ref{eq2-12}) within errors of the numerical results.
This deviation in small $N_{\mathrm{DW}}\xi$ might be caused by the finiteness of the simulation box which promotes walls
to collapse around the final time of the simulation (we observed a similar effect in our previous numerical study~\cite{2010JCAP...05..032H}).
In order to resolve this effect, we have to run simulations with higher spatial resolution (or larger simulation box), which is beyond the scope of this paper.
Here we just use the results with $N_{\mathrm{DW}}\xi\ge0.001$ to fit the data into the relation which we expect from the naive estimation (\ref{eq2-12}),
\begin{equation}
\tau_{\mathrm{dec}}\eta = \frac{\kappa}{\sqrt{N_{\mathrm{DW}}\xi}}\left(\frac{m}{0.1\eta}\right)^{1/2}. \label{eq3-2-4}
\end{equation}
We assume that $t_{\mathrm{dec}}$ is proportional to $m$ as we see in eq.~(\ref{eq2-12}) and write
this dependence explicitly in eq.~(\ref{eq3-2-4}). We discuss this point in the next subsection.
The best fit value for the numerical coefficient $\kappa$ turns out to be
\begin{equation}
\kappa = 2.70\pm0.03. \label{eq3-2-5}
\end{equation}
From eqs.~(\ref{eq3-2-4}) and (\ref{eq3-2-5}), we get the formula for the decay time of domain walls
\begin{equation}
\tau_{\mathrm{dec}} \simeq 8.5\times \sqrt{\frac{m}{N_{\mathrm{DW}}\xi\eta^3}}, \label{eq3-2-6}
\end{equation}
or in cosmic time,
\begin{equation}
t_{\mathrm{dec}} \simeq 18\times\left(\frac{m}{N_{\mathrm{DW}}\xi\eta^2}\right). \label{eq3-2-7}
\end{equation}
We conclude that the simple scaling of the decay time with the bias parameter is robust at least in the parameter range where the finite box effect can be neglected
and agrees with simple analytic expectation given by eq.~(\ref{eq2-12}), except the numerical factor of $\cal O$(10).
We will extrapolate this relation for much smaller values of $\xi$ in the next section.

\subsubsection{\label{sec3-2-2}Dependence on $m$}
As we mentioned before, we can not make the value of $m/\eta$ arbitrarily small since we invalidate our assumption about the approximate discrete symmetry.
Furthermore, $m/\eta$ can not be so large because the core size of the string $\delta_s$ must be smaller than the width of the domain wall $\delta_w$ for the formation of
stable networks~\cite{1999PhRvD..59b3505C}: $\delta_s/\delta_w\simeq m/(\eta\sqrt{\lambda}) \ll 1$ (for the parameters we used in the simulations, $m/(\eta\sqrt{\lambda}) \approx 0.32$).
Therefore, the range of $m/\eta$ which we can choose in the actual numerical simulation is narrow.

\begin{figure}[htbp]
\begin{center}
\includegraphics{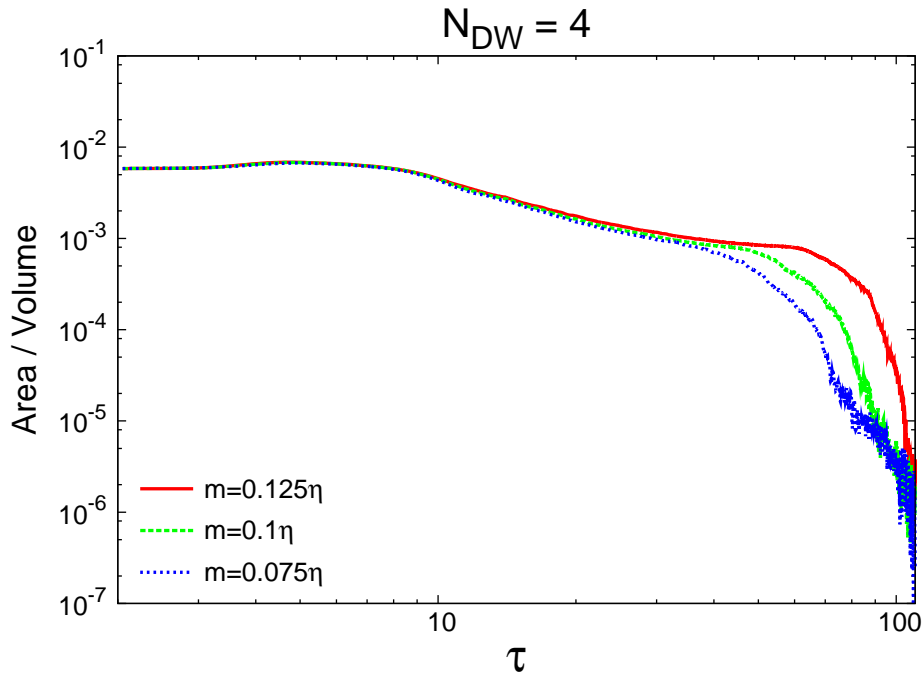}
\end{center}
\caption{The time evolution of the comoving area density of domain walls for various values of $m$. We set other parameters as $\lambda=0.1$, $N_{\mathrm{DW}}=2$, and $\xi=0.0004$.}
\label{fig6}
\end{figure}

Figure~\ref{fig6} shows the $m/\eta$ dependence of the time evolution of the comoving area density of domain walls. Although the range of $m/\eta$
is limited, from this figure we can see that the decay time of domain walls (in conformal time) is proportional to $\sqrt{m}$, as we anticipated in eq.~(\ref{eq2-12}).
Therefore, we assume that this dependence on $m$ is correct for other value of $m/\eta$, and use the relation (\ref{eq3-2-7}) for
the evaluation of the decay time of domain walls.

\section{\label{sec4}Observational constraints}
In the previous section, we treated $\xi$ as a free parameter except the condition that it must be sufficiently small.
However, actual observations can put stringent constraints on this parameter.
The purpose of this section is to give a quantitative constraints on the parameters by using the numerical result~(\ref{eq3-2-7}).

\subsection{\label{sec4-1}NEDM bound}
The first constraint comes from the fact that the existence of the asymmetric term $\delta V$ spoils the PQ solution
to the CP violation problem in QCD~\cite{1992PhLB..282..137K, 1992PhLB..282..132H, 1992PhRvD..46..539B, 1997PhRvD..55.5826D}.
The QCD Lagrangian contains a term
\begin{equation}
{\cal L}_{\theta} = \frac{\bar{\theta}}{32\pi^2}G^{a\mu\nu}\tilde{G}^a_{\ \mu\nu}, \label{eq4-1-1}
\end{equation}
where $G^{a\mu\nu}$ is the gluon field strength, $\tilde{G}^a_{\ \mu\nu}\equiv\frac{1}{2}\epsilon_{\mu\nu\lambda\sigma}G^{a\lambda\sigma}$
is the dual of it, and $\bar{\theta}$ is an dimensionless parameter. This term violates both P and CP and leads to an neutron electric dipole moment (NEDM)
with large amplitude which conflict with the experiments. The original PQ idea is to take $\bar{\theta}$ as a dynamical field with a potential minimized at $\bar{\theta}=0$.
However, if we introduce the asymmetric term $\delta V$, it shifts the minimum of the potential from $\bar{\theta}=0$.

Let us define axion field $a$ as
$\phi=\eta e^{ia/\eta}$. Substituting it into eq.~(\ref{eq2-2}) and expanding for small $a/\eta$, we can derive the effective potential for axion field~\cite{1992PhRvD..46..539B}
\begin{equation}
V(a) \simeq \frac{1}{2}m^2_{\mathrm{phys}}a^2-m^2_{\mathrm{phys}}\eta a\bar{\theta}/N_{\mathrm{DW}}, \label{eq4-1-2}
\end{equation}
where $m_{\mathrm{phys}}$ is the effective mass of the axion
\begin{equation}
m^2_{\mathrm{phys}} = m^2 + 2\xi\eta^2\cos\delta, \label{eq4-1-3}
\end{equation}
and $\bar{\theta}$ is the minimum of the potential
\begin{equation}
\bar{\theta} = N_{\mathrm{DW}}\left\langle\frac{a}{\eta}\right\rangle = \frac{2N_{\mathrm{DW}}\xi\eta^2\sin\delta}{m^2 + 2\xi\eta^2\cos\delta}. \label{eq4-1-4}
\end{equation}
The recent upper bound on $\bar{\theta}$ from the experiments is $\bar{\theta}<0.7\times 10^{-11}$~\cite{2006PhRvL..97m1801B, 2010RvMP...82..557K}. 
Assuming $\xi \ll m^2/\eta^2$ and $\sin\delta\approx 1$, we have the condition
\begin{equation}
\bar{\theta}\approx \frac{2N_{\mathrm{DW}}\xi\eta^2}{m^2} < 0.7\times 10^{-11}. \label{eq4-1-5}
\end{equation}
This gives an upper limit on $\xi$.

\subsection{\label{sec4-2}Overclosure bound}
If $\xi$ is sufficiently small, string-wall networks continue to exist for a long time.
Since they evolve to maintain the scaling property as eq.~(\ref{eq2-8}), their energy density decreases slower than that of matter and radiation,
and they eventually overclose the universe. If it occurs, they cause the fast expansion and affect the thermal history of the universe.
Therefore, another constraint comes from the condition that domain walls must not dominate the energy density of the universe.

Let us estimate the time at which domain walls dominate the energy density of the universe. Assuming the radiation dominated era, the energy density of the universe
and domain walls (at the cosmic time $t$) are respectively $\rho_c(t)=(3/8\pi G)H^2=3/32\pi G t^2$ and $\rho_{\mathrm{wall}}\simeq \sigma/H^{-1}=\sigma/2t$,
where $G$ is the Newton's constant and $\sigma$ is given by  eq.~(\ref{eq2-6}). By equating these two, we find that the wall domination occurs at the time
\begin{equation}
t_{\mathrm{WD}} = \frac{3}{16\pi G\sigma}. \label{eq4-2-1}
\end{equation}
We must require $t_{\mathrm{dec}}<t_{\mathrm{DW}}$, from which we obtain the condition
\begin{equation}
\xi > 3\times 10^3 \times N_{\mathrm{DW}}^{-3}\left(\frac{m}{M_P}\right)^2, \label{eq4-2-2}
\end{equation}
where we used eq.~(\ref{eq3-2-7}) for $t_{\mathrm{dec}}$, and $M_P$ is the Planck mass. Eq.~(\ref{eq4-2-2}) gives a lower bound on $\xi$.

There is a subtlety in this discussion. To make it clear, let us calculate the temperature at which the wall domination occurs.
By using the relation $H_{\mathrm{WD}}^2=(8\pi^3g_{\mathrm{WD}}/90M_P^2)T_{\mathrm{WD}}^4=1/(2t_{\mathrm{WD}})^2$, where
$g_{\mathrm{WD}}$, $T_{\mathrm{WD}}$ and $H_{\mathrm{WD}}$ are the relativistic degrees of freedom, temperature and Hubble parameter at the time $t_{\mathrm{WD}}$
respectively, we found
\begin{equation}
T_{\mathrm{WD}} = 8\times 10^{-2}\times \left(\frac{10}{g_{\mathrm{WD}}}\right)^{1/4}\left(\frac{F_a}{10^{12}\mathrm{GeV}}\right)^{1/2}\mathrm{MeV}. \label{eq4-2-3}
\end{equation}
From this equation, we see that the wall domination occurs after the time of big bang nucleosynthesis (BBN) which occurs around the temperature $T_{\mathrm{BBN}}\simeq 1$MeV.
Furthermore, the upper bound on $\xi$ given by eq.~(\ref{eq4-1-5}) implies that walls must survive until the BBN epoch, as we will see below (in figure~\ref{fig7}).
Therefore, in this scenario string-wall networks exist during the BBN epoch, and we must consider whether they spoil the success of the standard BBN scenario.

If the energy density of domain walls becomes sufficiently large during the BBN epoch, it alter the expansion rate of the universe, which
affects the helium abundance observed today.
The energy density of domain walls contributes as an extra energy component to the total energy density of the universe during the BBN epoch,
which is usually parameterized as an effective number of neutrino species $N_{\nu}$ (see e.g.~\cite{2000PhR...331..283M}):
\begin{equation}
\rho_{\mathrm{extra}}(t_{\mathrm{BBN}}) = \frac{\pi^2}{30}\frac{7}{8}(N_{\nu}-3)T_{\mathrm{BBN}}^4. \label{eq4-2-4}
\end{equation}
Requiring $\rho_{\mathrm{extra}}(t_{\mathrm{BBN}})=\rho_{\mathrm{wall}}(t_{\mathrm{BBN}})=\sigma H_{\mathrm{BBN}}$, where $H_{\mathrm{BBN}}$ is the Hubble
parameter at the BBN epoch, we found
\begin{equation}
N_{\nu}-3 = 5.8\times \frac{g_{\mathrm{BBN}}^{1/2}\sigma}{M_PT_{\mathrm{BBN}}^2} = 8.4\times10^{-2}\times\left(\frac{F_a}{10^{12}\mathrm{GeV}}\right), \label{eq4-2-5}
\end{equation}
where $g_{\mathrm{BBN}}=10.75$ is the relativistic degree of freedom at the BBN epoch.
The recent observations indicate the value 3.68-3.80 for $N_{\nu}$, but there are various systematic uncertainties~\cite{2010ApJ...710L..67I}.
This implies that the extra contribution to the energy density at the BBN epoch must be at most $N_{\nu}\lesssim4$. From eq.~(\ref{eq4-2-5})
we can see that this limit is satisfied for all values of $F_a$ which do not exceed the cosmological bound [i.e. eq.~(\ref{eq4-4-4})].
Therefore, domain walls do not have a significant effect to the expansion rate during the BBN epoch.

\subsection{\label{sec4-3}Cold axions from the decay of domain walls}
Domain walls would decay after the BBN epoch and 
the energy stored in walls is converted mainly into the radiation of axion particles and gravitational waves.
Let us comment on the effect of the axions produced by domain walls (we will discuss about gravitational waves in the next section).
These axions are barely relativistic~\cite{1999PhRvD..59b3505C, 1994PhRvD..50.4821N} and have a Compton length much
larger than that of the light elements produced in the BBN epoch. Therefore they can not spoil the light element abundance observed today.

However, since these axions are barely relativistic, they contribute to the energy density of the universe as a cold matter component.
The energy density of cold axions produced by domain walls must not exceed the energy density of the radiations until the time of 
equality between matter and radiation $t_{\mathrm{eq}}$, which might give a severe constraint for this scenario.

Let us estimate the abundance of cold axions at the time $t_{\mathrm{eq}}$. Suppose that the fraction $r$ of the energy stored in the wall
become barely relativistic axions at the time of decay of domain walls. Then, the energy density of cold axions at the time of decay is given by
\begin{equation}
\rho_a(t_{\mathrm{dec}}) = r\rho_{\mathrm{wall}}(t_{\mathrm{dec}}), \label{eq4-3-1}
\end{equation}
where $\rho_{\mathrm{wall}}(t_{\mathrm{dec}}) = \sigma/2t_{\mathrm{dec}}$ is the energy density of domain walls at the time $t_{\mathrm{dec}}$.
The number density of cold axions at the time $t>t_{\mathrm{dec}}$ can be estimated as
\begin{equation}
n_a(t) = \frac{\rho_a(t_{\mathrm{dec}})}{\omega_a}\left(\frac{a(t_{\mathrm{dec}})}{a(t)}\right)^3, \label{eq4-3-2}
\end{equation}
where $\omega_a=\gamma m(t_{\mathrm{dec}})$ is the average energy of the radiated axions, and $m(t_{\mathrm{dec}})$ is the axion mass at the time $t_{\mathrm{dec}}$.
$\gamma$ is the Lorentz factor, and the past numerical study indicates $\gamma\simeq 60$~\cite{1999PhRvD..59b3505C}. Using eq.~(\ref{eq4-3-2}),
we find the fraction of the energy density of cold axions at the time $t_{\mathrm{eq}}$ as
\begin{equation}
\Omega_a(t_{\mathrm{eq}}) = \frac{m(t_{\mathrm{eq}})n_a(t_{\mathrm{eq}})}{\rho_c(t_{\mathrm{eq}})} \approx \frac{r\sigma}{2\gamma t_{\mathrm{eq}}\rho_c(t_{\mathrm{eq}})}\left(\frac{a(t_{\mathrm{dec}})}{a(t_{\mathrm{eq}})}\right)^3, \label{eq4-3-3}
\end{equation}
where $m(t_{\mathrm{eq}})$ is the axion mass at the time $t_{\mathrm{eq}}$, and $\rho_c(t_{\mathrm{eq}})$ is the critical energy density of the universe
at the time $t_{\mathrm{eq}}$. We used the approximation $m(t_{\mathrm{eq}})/m(t_{\mathrm{dec}})\approx 1$ at the second equality. Substituting eqs.~(\ref{eq2-6})
and (\ref{eq3-2-7}) for $\sigma$ and $t_{\mathrm{dec}}$, respectively, and using $\rho_c(t_{\mathrm{eq}}) = 2.27\times 10^{-33}(\Omega_Mh^2)^4\mathrm{GeV^4}$
(see, e.g.~\cite{2008cosm.book.....W}) and $a(t_{\mathrm{dec}})/a(t_{\mathrm{eq}})=(\sqrt{2}t_{\mathrm{dec}}H_{\mathrm{eq}})^{1/2}$,
where $\Omega_M$ is the ratio of the total matter density today to the critical density, $h$ is the current value of the Hubble parameter in units of 100 km sec$^{-1}$Mpc$^{-1}$
and $H_{\mathrm{eq}}$ is the Hubble parameter at the time $t_{\mathrm{eq}}$, we obtain
\begin{equation}
\Omega_a(t_{\mathrm{eq}}) \approx 3\times 10^{-29}\times r\xi^{-1/2}N_{\mathrm{DW}}^{-3/2}\left(\frac{60}{\gamma}\right)\left(\frac{0.15}{\Omega_Mh^2}\right)\left(\frac{10^{12}\mathrm{GeV}}{F_a}\right)^{1/2}. \label{eq4-3-4}
\end{equation}

Requiring $\Omega_a(t_{\mathrm{eq}}) < \frac{1}{2}$, we found the condition
\begin{equation}
\xi > 3\times 10^{-57}\times r^2N_{\mathrm{DW}}^{-3}\left(\frac{60}{\gamma}\right)^2\left(\frac{0.15}{\Omega_Mh^2}\right)^2\left(\frac{10^{12}\mathrm{GeV}}{F_a}\right). \label{eq4-3-5}
\end{equation}
It gives another lower bound on $\xi$. Note that it depends on the value of $r$, the fraction of the wall energy converted into cold axions.
If $r$ is as large as ${\cal O}(0.1)$, the constraint~(\ref{eq4-3-5}) becomes much severer than that of eq.~(\ref{eq4-2-2}).
Furthermore, it depends on the average Lorentz factor $\gamma$ of the radiated axions. The computer simulations performed in~\cite{1999PhRvD..59b3505C}
found $\gamma\simeq 7$ when $\sqrt{\lambda}\eta/m\simeq 20$. Extrapolating this result into the realistic value $\sqrt{\lambda}\eta/m\sim {\cal O}(10^{26})$,
they obtained $\gamma\simeq 60$. However, it is not clear whether this numerical result persists up to the different scale of order $10^{26}$.
If the value of $\gamma$ is actually ${\cal O}(1)$ when $\sqrt{\lambda}\eta/m\sim {\cal O}(10^{26})$, the bound given by eq.~(\ref{eq4-3-5})
becomes much severer, and it may completely exclude the scenario which we considering here unless $r$ is considerably suppressed.

\subsection{\label{sec4-4}Axion window for a model with $N_{\mathrm{DW}}>1$}
The constraints (\ref{eq4-1-5}) and (\ref{eq4-2-2}) also depend on the parameter $m$, which is the axion mass given by
\begin{equation}
m \simeq 6\times 10^{-6}\mathrm{eV}\left(\frac{10^{12}\mathrm{GeV}}{F_a}\right), \label{eq4-4-1}
\end{equation}
where $F_a\equiv \eta/N_{\mathrm{DW}}$ is the axion decay constant~\cite{2010RvMP...82..557K}.
The value of $F_a$ has been constrained from various astrophysical and cosmological observations.
Here we just enumerate axion constraints obtained so far.

Most astrophysical searches give constraints on the coupling parameters between axions and other particles.
Axions interact with photons due to the coupling
\begin{equation}
{\cal L}_{a\gamma\gamma} = \frac{g_{a\gamma\gamma}}{4}aF^{\mu\nu}\tilde{F}_{\mu\nu}, \label{eq4-4-2}
\end{equation}
where $F^{\mu\nu}$ is the photon field strength, $\tilde{F}_{\mu\nu}\equiv \frac{1}{2}\epsilon_{\mu\nu\lambda\sigma}F^{\lambda\sigma}$ is the dual of it, 
$g_{a\gamma\gamma}\equiv \frac{\alpha}{2\pi F_a}c_{a\gamma\gamma}$ is the axion-photon coupling, and $\alpha=e^2/4\pi$ is the fine structure constant.
The numerical coefficient $c_{a\gamma\gamma}$ is determined by electromagnetic and color anomaly and the up/down
quark mass ratio $z\equiv m_u/m_d$ (see e.g.~\cite{2010RvMP...82..557K}). Usually it takes a value of ${\cal O}(1)$,
depending on models. Axions also have interactions with fermions
\begin{equation}
{\cal L}_{ajj} = -i\frac{C_jm_j}{F_a}a\bar{\psi}_j\gamma_5\psi_j, \label{eq4-4-3}
\end{equation}
where $\psi_j$ is the fermion field, $m_j$ is its mass, and $C_j$ is a numerical coefficient whose value depends on models.
In hadronic axion models such as KSVZ model~\cite{1979PhRvL..43..103K, 1980NuPhB.166..493S}, axions do not have interactions
with ordinary leptons and quarks. On the other hand, in DFSZ model~\cite{1981PhLB..104..199D, Zhitnitsky1980}, axions interact with
electrons so that $C_e=\cos^2\beta/3$, where $\tan\beta$ is the ratio of two Higgs vacuum expectation values.

The axion emission in the sun gives some constraints on the axion-photon coupling.
The energy loss by solar axion emission requires enhanced nuclear burning and increases solar $^8$B neutrino flux. The observation
of $^8$B neutrino flux gives a bound $g_{a\gamma\gamma}\lesssim 5\times 10^{-10}$GeV$^{-1}$
or $F_a/c_{a\gamma\gamma}\gtrsim 2.3\times 10^6$GeV~\cite{1999APh....10..353S,2008LNP...741...51R}. Also, axion helioscope experiments
give the bounds on the axion-photon coupling. For example, the recent result of the CERN Axion Solar Telescope (CAST)~\cite{2007JCAP...04..010A}
gives a bound $g_{a\gamma\gamma}<8.8\times 10^{-11}$GeV$^{-1}$, which corresponds to $F_a/c_{a\gamma\gamma}> 1.3\times 10^7$GeV.

The axion emission from the globular clusters gives another bound on $F_a$. From the helium-burning lifetimes of horizontal branch stars,
the axion-photon coupling can be constrained as $g_{a\gamma\gamma}\lesssim0.6\times10^{-10}$GeV$^{-1}$~\cite{1999ARNPS..49..163R},
which corresponds to $F_a/c_{a\gamma\gamma}> 2\times 10^7$GeV. On the other hand, the delay of helium ignition in low-mass red-giant branch stars
gives a limit on the axion-electron coupling, $\alpha_{aee}<0.5\times 10^{-26}$ or $g_{aee}<3\times10^{-13}$~\cite{1995PhRvD..51.1495R},
where $\alpha_{aee}\equiv g^2_{aee}/4\pi$, $g_{aee}\equiv C_em_e/F_a$ and $m_e$ is the mass of electrons.
For DFSZ model, it corresponds to $F_a/\cos^2\beta>8\times10^8$GeV.

The observation of white-dwarfs also gives the bounds on the axion-electron coupling. The observed luminosity of white-dwarfs
indicates $\alpha_{aee}\lesssim 1\times10^{-26}$~\cite{1986PhLB..166..402R}, which corresponds to $F_a/\cos^2\beta\gtrsim5\times10^8$GeV for DFSZ model.
Furthermore, white-dwarfs which show periodic variations in their light curves (known as ZZ-Ceti stars)
give the more restrictive bound on the coupling $\alpha_{aee}$. From the observed rate of change of the period of the star G117-B15A,
it was obtained that $\alpha_{aee}<1.3\times 10^{-27}$ or $g_{aee}<1.3\times10^{-13}$~\cite{2001NewA....6..197C},
which corresponds to $F_a/\cos^2\beta>1.3\times10^9$GeV for DFSZ model.

Finally, from the energy loss rate of the supernova (SN) 1987A~\cite{1990PhR...198....1R}, one can constrain the axion-nucleon coupling.
For a small value of the coupling, the mean free path of axions becomes larger than the size of the SN core (so called the ``free streaming" regime).
In this regime, the energy loss rate is proportional to the axion-nucleon coupling squared, and one can obtain the limit $F_a\gtrsim 4\times10^8$GeV~\cite{2008LNP...741...51R}.
On the other hand, for a large value of the coupling, axions are ``trapped" inside the SN core. In this regime, by requiring that the axion emission should not
have a significant effect on the neutrino burst, one can obtain another bound $F_a\lesssim {\cal O}(1)\times 10^6$GeV~\cite{1990PhRvD..42.3297B}.
However, in this ``trapped" regime, it was argued that the strongly-coupled axions with $F_a \lesssim {\cal O}(1)\times 10^5$GeV would have produced
an unacceptably large signal at the Kamiokande detector, and hence they are ruled out ~\cite{1990PhRvL..65..960E}.

Note that, there is a parameter region around $F_a \sim 10^6$GeV that cannot be excluded by the SN observation (so called the ``hadronic axion window")~\cite{1993PhLB..316...51C}.
In particular, in KSVZ model, it is possible that the axion-photon coupling vanishes due to the uncertainty of the quark mass ratio $m_u/m_d$ and
the bound on $g_{a\gamma\gamma}$ described above does not significantly constrain the value of $F_a$ \cite{1993PhLB..316...51C,1998PhLB..440...69M,2007JCAP...07..012B}.
In this case, we have to take the hadronic window into account, but it might be highly model dependent.
In the following, we only take a constraint $F_a\gtrsim 4\times10^8$GeV obtained in the ``free streaming" regime,
since it gives a strong constraint on $F_a$ both for KSVZ model and for DFSZ model.
We take it as a lower bound on $F_a$ and neglect other possible model dependences.
We note that this lower bound suffers from various uncertainties
such as the uncertainty of the axion-nucleon couplings and the expression for spin-density structure function
which is used to calculate the axion emission rate~\cite{2008LNP...741...51R}. These uncertainties may modify the lower bound on $F_a$
by a factor of ${\cal O}(1)$.

In addition to the astrophysical bounds described above, we must consider the cosmological bound on $F_a$. 
After the QCD phase transition, the axion field behaves like the cold dark matter (CDM) component of the universe
due to the so called misalignment mechanism~\cite{2010PhRvD..82l3508W}. It is required that the axion CDM abundance
must not exceed the abundance of CDM observed today. This cosmological requirement gives an upper bound on $F_a$.
If we include the contribution of axions radiated by string decay, the constraint becomes $F_a<3\times10^{11}$GeV~\cite{2010arXiv1012.5502H}.

\begin{figure}[htbp]
\begin{center}
\includegraphics[scale=0.45]{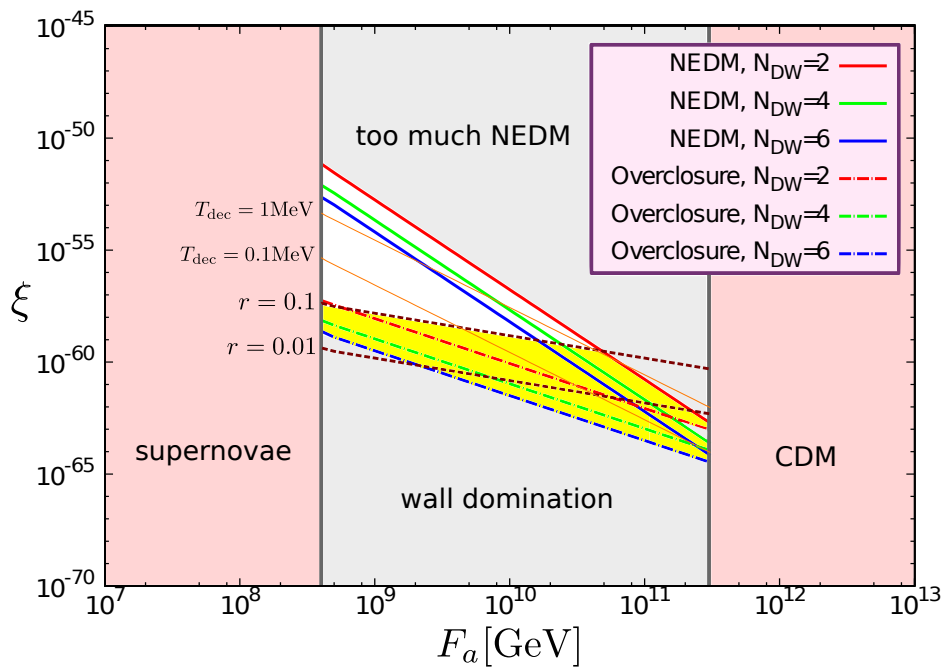}
\end{center}
\caption{The contours of the various observational constraints in the parameter space of $F_a$ and $\xi$. The solid line shows the bound given by eq.~(\ref{eq4-1-5}) and the dot-dashed
line shows the bound given by eq.~(\ref{eq4-2-2}), for a domain wall number $N_{\mathrm{DW}}=2$ (red), $N_{\mathrm{DW}}=4$ (green) and $N_{\mathrm{DW}}=6$ (blue), respectively.
The gray region denoted by ``too much NEDM" is the parameter space which induces too much CP violation
conflicting with the experiment of the NEDM, and the region denoted by ``wall domination"
is the parameter space in which the energy density of domain walls dominates the total energy density of the universe. These regions are observationally ruled out.
The vertical line shows the usual bound for the axion decay constant, given by eq.~(\ref{eq4-4-4}).
The pink region denoted by ``supernovae" is ruled out by the observation of the SN1987A~\cite{2008LNP...741...51R}, and the region denoted by
``CDM" is excluded by the condition that the axion CDM abundance exceeds the CDM abundance observed today.
Note that, there are many other astrophysical constraints and various model dependencies (see the text for details).
The thin solid lines represent the parameters in which $T_{\mathrm{dec}}=1$MeV and 0.1MeV,
where $T_{\mathrm{dec}}$ is the temperature at which the domain walls decay (we fixed $N_{\mathrm{DW}}=6$ for the evaluation of $T_{\mathrm{dec}}$).
The dotted lines show the bound given by eq.~(\ref{eq4-3-5}) for $r=0.1$ and $r=0.01$, where we fixed $\gamma=60$, $N_{\mathrm{DW}}=6$ and $\Omega_Mh^2=0.15$.
The yellow region might be excluded if $r$ is grater than $0.1$.}
\label{fig7}
\end{figure}

In summary, we take the astrophysical and cosmological bound on $F_a$ as
\begin{equation}
4\times 10^8\mathrm{GeV} < F_a < 3\times 10^{11}\mathrm{GeV}. \label{eq4-4-4}
\end{equation}
Combining this condition with previous constraints (\ref{eq4-1-5}), (\ref{eq4-2-2}) and (\ref{eq4-3-5}), we can draw the allowed region in the $\xi$-$F_a$ plane as shown in figure \ref{fig7}.
From this figure we can see that the allowed region is quite narrow, though it is not completely ruled out. We note that there is an unknown factor $r$.
If $r$ is not suppressed, the constraint for the model with $N_{\mathrm{DW}}>1$ would become much severe, as we see in figure~\ref{fig7}.
We also note that, as we described above, the axion-photon coupling may vanish for a certain value of $m_u/m_d$ in KSVZ model,
and the constraint space significantly opens up. However, this possibility is not shown in figure~\ref{fig7}.

\section{\label{sec5}Gravitational waves from axionic domain walls}
If domain walls survive for a long time, they produce gravitational waves. According to the recent numerical study~\cite{2010JCAP...05..032H},
gravitational waves produced by domain walls have almost flat spectrum from the frequency corresponding to the Hubble radius at $t_{\mathrm{dec}}$ to the
frequency corresponding to the inverse of the wall width, with an amplitude determined by a simple dimensional analysis. 

Let us estimate the amplitude and the frequency of gravitational waves produced by axionic domain walls.
The amplitude of gravitational waves observed today is characterized by a dimensionless quantity $\Omega_{\mathrm{gw}}$ defined by~\cite{2000PhR...331..283M}
\begin{equation}
\Omega_{\mathrm{gw}}(t_0) \equiv \frac{1}{\rho_c(t_0)} \frac{d\rho_{\mathrm{gw}}(t_0)}{d\log f}, \label{eq5-1}
\end{equation}
where $\rho_c(t_0)$ is the critical energy density of the universe today, $\rho_{\mathrm{gw}}(t_0)$ is the energy density of gravitational waves at the present time,
and $f$ is their frequency.
If the gravitational waves are produced at time $t_*$ (assuming the radiation dominated era), the amplitude observed today can be written as~\cite{1994PhRvD..49.2837K}
\begin{equation}
\Omega_{\mathrm{gw}}(t_0)h^2 = 3.60\times10^{-5}\left(\frac{10}{g_*}\right)^{1/3}\Omega_{\mathrm{gw}}(t_*), \label{eq5-2}
\end{equation}
where $g_*$ is the number of radiation degrees of freedom at the time $t_*$, 
and $\Omega_{\mathrm{gw}}(t_*) = \rho_{\mathrm{gw}}(t_*)/\rho_c(t_*)$. From a dimensional analysis, we obtain~\cite{2010JCAP...05..032H} $\rho_{\mathrm{gw}}(t_*)\sim G\sigma^2$,
where $\sigma$ is given by  eq.~(\ref{eq2-6}). By using $\rho_c(t_*) = 3/32\pi Gt_*^2$ and $t_*\simeq t_{\mathrm{dec}}$, where $t_{\mathrm{dec}}$
is given by eq.~(\ref{eq3-2-7}), we obtain
\begin{equation}
\Omega_{\mathrm{gw}}(t_0)h^2 \simeq 2\times 10^{-16}\times N_{\mathrm{DW}}^{-6}\left(\frac{10^{-58}}{\xi}\right)^2\left(\frac{10}{g_*}\right)^{1/3}\left(\frac{10^{12}\mathrm{GeV}}{F_a}\right)^4. \label{eq5-3}
\end{equation}

The frequency of gravitational waves redshifts as the universe expands. Therefore, the frequency observed today is estimated as
\begin{equation}
f(t_0) = \left(\frac{g_0}{g_*}\right)^{1/3}\frac{T_0}{T_*}f_*, \label{eq5-4}
\end{equation}
where $f_*$ is the frequency of
gravitational waves at the time $t_*$, $T_0=$2.725K is the temperature of the universe observed today, $T_*$ is the temperature at the time $t_*$,
and $g_0=3.36$ is the number of radiation degrees of freedom today. The numerical result obtained in~\cite{2010JCAP...05..032H}
implies that the frequency of gravitational waves ranges from the Hubble radius at the time $t_*$ ($f_*\sim H_*$) to the inverse of the wall width ($f_*\sim \delta^{-1}_w\simeq m$).
The frequency corresponding to the Hubble radius becomes
\begin{equation}
f_h(t_0) \simeq 3\times 10^{-9}\times N^{3/2}_{\mathrm{DW}}\left(\frac{\xi}{10^{-58}}\right)^{1/2}\left(\frac{10}{g_*}\right)^{1/12}\left(\frac{F_a}{10^{12}\mathrm{GeV}}\right)^{3/2} \mathrm{Hz}, \label{eq5-5}
\end{equation}
where we used the relation
\begin{equation}
H^2_* = \frac{8\pi^3g_*T^4_*}{90M_P^2} \simeq 1/(2t_{\mathrm{dec}})^2 \nonumber
\end{equation}
to eliminate $T_*$. Similarly, the frequency corresponding to the wall width becomes
\begin{equation}
f_w(t_0) \simeq 5\times 10^{-2}\times N^{-3/2}_{\mathrm{DW}}\left(\frac{10^{-58}}{\xi}\right)^{1/2}\left(\frac{10}{g_*}\right)^{1/12}\left(\frac{10^{12}\mathrm{GeV}}{F_a}\right)^{5/2} \mathrm{Hz}. \label{eq5-6}
\end{equation}

For example, if we take the parameters $N_{\mathrm{DW}}=4$, $F_a=10^{10}$GeV, $g_*=10$, and $\xi=10^{-58}$,
which are not ruled out by the observations described in the previous section,
the amplitude and the frequency of gravitational waves become $\Omega_{\mathrm{gw}}(t_0)h^2\simeq 5\times 10^{-12}$, $f_h\simeq 2\times 10^{-11}$Hz,
and $f_w\simeq 6\times 10^2$Hz. We note that this signal has an amplitude relevant to the future gravitational wave experiments. The ground-based experiments
are sensitive to the frequency around ${\cal O}$(10-10$^4$)Hz, which is corresponding to the frequency $f_w$ described above. As a ongoing ground-based experiment,
LIGO~\cite{1992Sci...256..325A} has sensitivities $\Omega_{\mathrm{gw}}h^2>10^{-6}$ today, and it is planned to upgrade to Advanced LIGO~\cite{AdvLIGO},
which would have sensitivity ${\cal O}$(10$^{-9}$). The LCGT~\cite{2002CQGra..19.1237K} is also planned with sensitivity similar to that of Advanced LIGO.
The Einstein Telescope~\cite{ET} would have more improved sensitivity of ${\cal O}$(10$^{-11}$).
Also, the space-borne interferometers such as LISA~\cite{LISA} and DECIGO~\cite{2006CQGra..23S.125K} are planned.
LISA is sensitive to the frequency around ${\cal O}$(10$^{-4}$-10$^{-1}$)Hz and it can reach sensitivity of ${\cal O}$(10$^{-11}$).
DECIGO is sensitive to the frequency band between LISA and ground-based interferometers,
and the ultimate sensitivity of DECIGO can reach ${\cal O}$(10$^{-20}$)~\cite{2001PhRvL..87v1103S}. Therefore, even if the ground-based experiments
do not have enough sensitivities, it is possible to detect some signal of gravitational waves in future space-borne interferometers.

We emphasize that the above argument is based on the numerical result with a simple model in which $Z_2$ symmetry is spontaneously broken.
Because there is another mass scale $\eta$ (the PQ scale) in addition to the two scales $m$ and $H_*$,
the precise shape of the spectrum of gravitational waves produced by string-wall networks
might be different from that we obtained numerically in~\cite{2010JCAP...05..032H}.
Therefore, it is necessary to perform three dimensional lattice simulation
to calculate the spectrum of gravitational waves produced by string-wall networks. Since it is computer intensive task, we leave it as a future work.

\section{\label{sec6}Conclusions}
In this paper, we investigated the evolution of string-wall networks which arise in axion models based on the two dimensional lattice simulations.
In addition to the usual cosine potential of the axion field, we added the bias term $\delta V$ which makes domain walls unstable.
We confirmed the result of the simulations performed by Ryden et al.~\cite{1990ApJ...357..293R}: Walls with $N_{\mathrm{DW}}=1$ quickly decay after the
beginning of the simulation, and walls with $N_{\mathrm{DW}}>1$ evolve maintaing the scaling property if we do not include the bias.
We also simulated the evolution of unstable domain walls which decay due to the existence of the asymmetric term $\delta V$ in the potential
and obtained the time scale of the decay as eq.~(\ref{eq3-2-7}). 
This time scale is consistent with the naive estimation (\ref{eq2-12}) except the numerical coefficient.
The observational bounds which come from
the NEDM and the overclosure of domain walls
can constrain the parameter of the bias $\xi$ together with the axion decay constant $F_a$ as eqs.~(\ref{eq4-1-5}) and (\ref{eq4-2-2}), and it is found that the
allowed region is rather narrow.
In addition to these constraints, we must require that the abundance of cold axions produced by the decay of domain walls
should not exceed the CDM abundance at the time of equality between matter and radiation. This gives another bound~(\ref{eq4-3-5}),
and it depends on the fraction $r$ of the wall energy converted into cold axions. This constraint might be severer than that obtained from the condition
of the overclosure of domain walls unless $r$ is considerably suppressed.
However, if such string-wall networks have existed, gravitational waves produced by domain walls would be detectable 
in future interferometers.

This issue provides a new observational method for axion physics in the sense that we might be able to probe axion models via gravitational wave astronomy.
If a signal of gravitational waves which we described in section~\ref{sec5} is detected in the future experiments,
it might be possible to measure the parameters such as a bias $\xi$ and the axion decay constant $F_a$ from the spectrum of gravitational waves.
Even if we can not detect any signal, we may constrain the window of the allowed region described in section~\ref{sec4-4} or completely rule out the model with $N_{\mathrm{DW}}>1$.

For now we estimated the amplitude and the frequency of gravitational waves based on the simple dimensional analysis,
but the precise shape of the spectrum of gravitational waves is not determined.
Especially, the spectrum of gravitational waves produced by string-wall networks which arise when $Z_{N_{\mathrm{DW}}}$ symmetry
is spontaneously broken might be different from that with the simple model in which $Z_2$ symmetry is spontaneously broken.
Also, the value of $r$, which is the fraction of the wall energy converted into cold axions when the decay of walls, is undetermined.
The constraint for the axion models with $N_{\mathrm{DW}}>1$ might become much severer if the value of $r$ is not so small.
These undetermined factors will be settled by direct numerical simulations, which we will tackle in the future work.

\begin{acknowledgments}
We would like to thank F. Takahashi for useful discussions. KS thank T. Sekiguchi and K. Miyamoto
for discussion on the axion abundance. Numerical computation in this work was carried out at the 
Yukawa Institute Computer Facility. 
This work is supported by Grant-in-Aid for
Scientific research from the Ministry of Education, Science, Sports, and
Culture (MEXT), Japan, No.14102004 and No.21111006 (M.K.)  and also by
World Premier International Research Center Initiative (WPI Initiative), MEXT, Japan.
KS is supported by the Japan Society for the Promotion of Science (JSPS) through research fellowships.
TH was supported by JSPS Grant-in-Aid for Scientific Research (A) No. 21244033.
\end{acknowledgments}


\appendix

\section{Notes on the numerical computations}
\label{secA}

In this appendix we describe the setup for our numerical computations.

\subsection{\label{secA1}Formulation}
We define two real scalar fields $\phi_1$ and $\phi_2$ such that $\phi = \phi_1 + i\phi_2$.
In the numerical studies, we normalize the dimensionful quantities in the unit of $\eta$. For example, $\phi\to\phi/\eta$, $t\to t\eta$, etc. 
With this normalization, the equations of motion for two real scalar fields derived from the Lagrangian (\ref{eq2-1}) with the FRW background are
\begin{eqnarray}
\ddot{\phi}_1 + 3H\dot{\phi}_1 - \frac{\nabla^2}{a^2(t)}\phi_1 &=& -\lambda\phi_1(\phi_1^2+\phi_2^2-1) + 2\xi\cos\delta + \frac{m^2}{N_{\mathrm{DW}}|\phi|}\sin\theta\sin N_{\mathrm{DW}}\theta, \label{eqA-1}\\
\ddot{\phi}_2 + 3H\dot{\phi}_2 - \frac{\nabla^2}{a^2(t)}\phi_2 &=& -\lambda\phi_2(\phi_1^2+\phi_2^2-1) + 2\xi\cos\delta - \frac{m^2}{N_{\mathrm{DW}}|\phi|}\cos\theta\sin N_{\mathrm{DW}}\theta, \label{eqA-2}
\end{eqnarray}
and the potential energy can be written in the form
\begin{equation}
V(\phi) = \frac{\lambda}{4}(|\phi|^2-1)^2 + \frac{m^2}{N_{\mathrm{DW}}^2}(1 - \cos N_{\mathrm{DW}}\theta) -2|\phi|\xi\cos(\theta -\delta) + 2\xi, \label{eqA-3}
\end{equation}
where we add a constant for $V(\phi)$ in order to make $V(\phi)=0$ at the minimum of the potential. 

In the numerical simulations, we use the conformal time $\tau$, and normalize $\phi$ by the scale factor,
\begin{eqnarray}
&&\frac{d}{dt} = \frac{1}{a}\frac{d}{d\tau},\quad \phi\equiv\frac{\bar{\phi}}{a},\label{eqA-4} \\
&&\ddot{\phi}+3H\dot{\phi} = \frac{1}{a^3}\left(\bar{\phi}''-\frac{a''}{a}\bar{\phi}\right), \label{eqA-5}
\end{eqnarray}
where a prime represents a derivative with respect to $\tau$. Note that, in the radiation dominated universe, $a''/a=0$ in the last equation.
Then, from eqs. (\ref{eqA-1}), (\ref{eqA-2}) and (\ref{eqA-5}), we obtain
\begin{eqnarray}
\bar{\phi}''_1 - \nabla^2\bar{\phi}_1 &=& -\lambda\bar{\phi}_1(|\bar{\phi}|^2-a^2) + 2a^3\xi\cos\delta + \frac{a^4m^2}{N_{\mathrm{DW}}|\bar{\phi}|}\sin\theta\sin N_{\mathrm{DW}}\theta, \label{eqA-6}\\
\bar{\phi}''_2 - \nabla^2\bar{\phi}_2 &=& -\lambda\bar{\phi}_2(|\bar{\phi}|^2-a^2) + 2a^3\xi\cos\delta - \frac{a^4m^2}{N_{\mathrm{DW}}|\bar{\phi}|}\cos\theta\sin N_{\mathrm{DW}}\theta. \label{eqA-7}
\end{eqnarray}
We choose the initial time of the simulation so that $t_i =1$ and $a(t_i)=1$.
This corresponds to the condition $\tau_i=2$ in conformal time.

The simulations are performed in the comoving box with size $b$ (in the unit of $\eta^{-1}$). The lattice spacing is $\delta x = b/N$ where $N$ is
the number of grid points (here we take $N=4096$). The physical scale of the Hubble radius $H^{-1}=2t$, the width of the wall $\delta_w=m^{-1}$,
and the core size of the string $\delta_s = 1/\sqrt{\lambda}$ divided by the physical lattice spacing $\delta x_{\mathrm{phys}}=a(t)\delta x$ are respectively
\begin{equation}
\frac{H^{-1}}{\delta x_{\mathrm{phys}}} = \frac{N\tau}{b},\qquad \frac{\delta_w}{\delta x_{\mathrm{phys}}} = \frac{N}{bm}\left(\frac{\tau_i}{\tau}\right), \quad \mathrm{and}\quad\frac{\delta_s}{\delta x_{\mathrm{phys}}} = \frac{N}{b\lambda^{1/2}}\left(\frac{\tau_i}{\tau}\right). \label{eqA-8}
\end{equation}
In our simulation we take $b=230$, $m=0.1$ and $\lambda=0.1$. Then, at the end of the simulation $\tau_f=110$, these ratios become
$H^{-1}/\delta x_{\mathrm{phys}}\simeq1959 < N$, $\delta_w/\delta x_{\mathrm{phys}}\simeq3.2$, and $\delta_s/\delta x_{\mathrm{phys}}\simeq1.02$.
Therefore, these length scales are marginally resolvable even at the final time of the simulation.

We put the periodic boundary condition in the configuration of the scalar fields. We solve the time evolution of the fields by using the fourth order
Runge-Kutta method.

\subsection{\label{secA2}Initial conditions}
The initial conditions are similar with that used in~\cite{2001PhRvD..64l3517F}. We treat $\phi_1$ and $\phi_2$ as two independent real scalar fields
so that each of them has quantum fluctuation at the initial time with correlation function in the momentum space given by
\begin{eqnarray}
\langle\phi_i({\bf k})\phi_i({\bf k}')\rangle &=& \frac{1}{2k}(2\pi)^3\delta^{(3)}({\bf k+k'}), \label{eqA-9}\\
\langle\dot{\phi}_i({\bf k})\dot{\phi}_i({\bf k}')\rangle &=& \frac{k}{2}(2\pi)^3\delta^{(3)}({\bf k+k'}). \qquad (i=1,2) \label{eqA-10}
\end{eqnarray}
Since the effective squared masses of the fields quickly become negative at the initial time (i.e. on the top of the mexican hat potential),
we used the massless fluctuations as the initial conditions, replacing the factor $\sqrt{k^2+m^2}$ with $k$ in the above formulae.
We also put the momentum cutoff $k_{\mathrm{cut}}$ above which all fluctuations are set to zero in order to eliminate the unphysical noise
which comes from high frequency modes in the field distributions. Here we set $k_{\mathrm{cut}}=1$.

Note that, although we describe the results of the computation on the ``two dimensional lattice" in the body of paper, 
the dimensionality of delta functions in eqs. (\ref{eqA-9}) and (\ref{eqA-10}) is 3.
We performed numerical simulations on the two dimensional lattice, which means that we choose the number of lattice point in $z$-axis to be $N=1$ while we take
$N=4096$ in $x$- and $y$- axes. This choice corresponds to a three dimensional field theory with plane symmetry along z-axis.

We generate initial conditions in momentum space as Gaussian random amplitudes satisfying eqs. (\ref{eqA-9}) and (\ref{eqA-10}),
then Fourier transform them into the configuration space to give the spatial distributions of the fields.
It is possible to start the simulation with more realistic condition, for example, with the thermal initial conditions.
Thermal initial conditions are relevant if we are interested in the formation of global strings,
since the formation time and initial spatial distribution of strings are affected by the initial amplitude given at high temperature.
After the string formation, the field distribution just before the QCD phase transition is determined by the vacuum 
configuration around global string networks. Then, the formation of axionic domain walls occurs if the expansion rate of the universe becomes less than the mass of axion.
Therefore, thermal initial conditions are not relevant in the QCD era. In any case, it is difficult to reproduce the whole process described above in the numerical simulations
because of the limitation of the dynamical range. However, since it seems that the scaling property is not so much affected by the initial field configurations and we are 
interested in the evolution of the fields after the formation of the string-wall networks, we expect that the results were qualitatively unchanged if we used the different initial 
conditions.

\subsection{\label{secA3}Calculation of the area density}
We used the similar algorithm for calculation of the area density of domain walls with that used in~\cite{2010JCAP...05..032H}.
We sum up the quantity $\delta_{\pm}$ which takes either 1 or 0 for each of the links (the neighboring grid points that differ by one) with the weighting factor as
\begin{equation}
A/V = C\sum_{\mathrm{links}}\delta_{\pm}\frac{|\nabla\theta|}{|\theta_{,x}| + |\theta_{,y}| + |\theta_{,z}|}, \label{eqA-11}
\end{equation}
where $\theta_{,x}$, etc. is a derivative of $\theta({\bf x})$ with respect to $x$, and $C$ is a coefficient chosen so that $A/V =1$ is satisfied when all the 
links have the value $\delta_{\pm}=1$. If $N_{\mathrm{DW}}=1$, we put $\delta_{\pm}=1$ at the location where $\phi_2$ has different signs and $\phi_1<0$ at the
two ends of the link (i.e. at the location on which $\theta=\pi$). If $N_{\mathrm{DW}}>1$, we divide the codomain of $\theta$ ($2\pi$) into $N_{\mathrm{DW}}$ domains,
and put $\delta_{\pm}=1$ at the location where $\theta$ belongs to different domains at the two ends of the link. Otherwise we put $\delta_{\pm}=0$.


\end{document}